\documentclass[11pt,a4paper]{article}
\usepackage{amsmath,amssymb,amsthm,eucal,cite,color}
\usepackage[top=3cm,right=3cm,left=2.5cm,bottom=3cm]{geometry}
\usepackage{framed,graphicx,multirow}

\title{On the spacetime structure of infrared divergencies in QED}

\author{Christian Ga\ss, Karl-Henning Rehren, Felix C. Tippner
\thanks{Email: 
\texttt{cgass@uni-goettingen.de,
  krehren@gwdg.de, felix.tippner@uni-goettingen.de}} \\[2mm] Institute for
Theoretical Physics \\ Georg-August University G\"ottingen \\ Friedrich-Hund-Platz 1,
37077 G\"ottingen, Germany}

\date{\today}

\newcommand{\be}{\begin{equation}}   
\newcommand{\ee}{\end{equation}} 

\newcommand{\eps}{\varepsilon}    
\newcommand{\wh}{\widehat}        
\newcommand{\wt}{\widetilde}      
\newcommand{\pa}{\partial}      

\DeclareMathOperator{\arcsinh}{arcsinh}      
\DeclareMathOperator{\IM}{Im}       

\def\wick#1{\mathopen:#1\mathclose:} 

\def\wickv#1{\wick{#1}_{v}}

\def\duo<#1,#2>{\langle#1,#2\rangle} 

\numberwithin{equation}{section}

\theoremstyle{plain}
\newtheorem{thm}{Theorem}[section]  
\newtheorem{prop}[thm]{Proposition} 
\newtheorem{lemma}[thm]{Lemma}       
\newtheorem{coro}[thm]{Corollary}   
\newtheorem{remk}[thm]{Remark}      

\theoremstyle{definition}

\newtheoremstyle{example}
   {\topsep}{\topsep}{\small}{0pt}%
   {\bfseries}{.}{ }{}
\theoremstyle{example}

\newtheoremstyle{exercise}
   {\topsep}{\topsep}{\small}{0pt}%
   {\bfseries}{.}{ }{}
\theoremstyle{exercise}

\theoremstyle{remark}

\DeclareRobustCommand{\qned}{\ifmmode
  \else \leavevmode\unskip\penalty9999 \hbox{}\nobreak\hfill \fi
  \quad\hbox{\qnedsymbol}}
\newcommand{\qnedsymbol}{$\boxminus$} 

\makeatletter
\renewcommand{\section}{\@startsection{section}{1}{\z@}%
                        {-3.5ex \@plus -1ex \@minus -.2ex}%
                        {2.3ex \@plus.2ex}%
                        {\normalfont\large\bfseries}}
\renewcommand{\subsection}{\@startsection{subsection}{2}{\z@}%
                        {-3.25ex \@plus -1ex \@minus -.2ex}%
                        {1.5ex \@plus .2ex}%
                        {\normalfont\normalsize\bfseries}}
\renewcommand{\subsubsection}{\@startsection{subsubsection}{3}{\z@}%
                        {-3.25ex \@plus -1ex \@minus -.2ex}%
                        {1.5ex \@plus .2ex}%
                        {\normalfont\normalsize\bfseries}}
\renewcommand{\@dotsep}{200} 
\makeatother

\def\bea#1{\begin{eqnarray}\label{#1}}
\def\eea{\end{eqnarray}}
\def\ba{\begin{array}}
\def\ea{\end{array}}
\def\bfr{\begin{framed}}
  \def\efr{\end{framed}}

\def\eref#1{Eq.~(\ref{#1})}         
\def\lref#1{Lemma~\ref{#1}}
\def\pref#1{Prop.~\ref{#1}}
\def\cref#1{Cor.~\ref{#1}}
\def\rref#1{Rem.~\ref{#1}}         
\def\sref#1{Sect.~\ref{#1}}
\def\aref#1{App.~\ref{#1}}
\def\erw#1{\langle #1\rangle}
\def\Erw#1{\big\langle #1\big\rangle}

\def\sign{{\rm sign}\,} 
\def\ioi{\int_0^\infty}  
\def\RR{\mathbb{R}}

\def\ol{\overline}
\def\sign{\mathrm{sign}}
\def\Li{\mathrm{Li}}
\def\lra{\leftrightarrow}
\def\blue#1{{\color{blue}#1}}
\def\bpm{\begin{pmatrix}}
  \def\epm{\end{pmatrix}}
\def\inv{^{-1}}\parskip2mm
\def\vv#1{\vert \vec #1\vert}

\def\edo{\end{document}}\parindent0mm

\begin{document}

\maketitle

\begin{abstract}
We investigate analytic properties of string-integrated
correlation functions and propagators with emphasis on their infrared
behaviour. These are relevant in various
models of quantum field theory with massless fields, including QED. 
\end{abstract}

\section{Motivation}
\label{s:intro}

A conceptually new approach to QED is presented in \cite{MRS3} (see
also \cite{S2,MRS2}).  It is designed to better understand the
long-distance behaviour of QED, including the uncountable superselection structure of charged
states due to their ``asymptotic photon clouds'', and the
infraparticle nature of the electron. The latter is manifest in a sharp lower end of the mass
spectrum with a singular set-on of the continuum due to the attached
soft photons. The new approach properly addresses and solves the problem with the quantum Gauss Law on
the physical Hilbert space: if the charged field were a local quantum
field, the integrated electric flux at spacelike infinity would commute
with it and cannot act
as the generator of the $U(1)$ symmetry. The tight relations among these physical
features have been known since long \cite{FPS,FMS,Bu2,Bu3}, while a
way to incorporate them into a model was so far lacking -- apart from a
simple model in 1+1 spacetime dimensions \cite{S1}.

The most prominent role in the new approach is played by an auxiliary quantum field formally defined as  
\bea{phiAK}\phi(x,e)= \int_{\Gamma_{x,e}} A^K_\mu(y)dy^\mu = \int_0^\infty ds\, A^K_\mu(x+se)e^\mu\eea
(introduced in more detail below; $\Gamma_{x,e}$ is a straight curve
from $x$ along a direction $e$ to infinity, and the superscript $K$ stands for ``Krein space''
emphasizing the indefinite metric of the usual Feynman gauge Fock
space). It is the main purpose of these notes to investigate details
of its infrared behaviour in position space.

The infrared superselection sectors of QED arise by exponentiating the
field $\phi(x,e)$, smeared with suitable functions
$c(e)$. Because \eref{phiAK} is infrared divergent as it stands, a suitable infrared cut-off function $v(k)$ is needed
that allows to extend the Fourier transform of the two-point function
to $k=0$ (as a distribution). With an appropriate regularization by a mass $m\to0$, the
correlation functions of exponentiated fields (``vertex
operators'')
\bea{wickv}N_v(c)\cdot \wick{e^{iq\phi(x,c)}}_{v}\eea
contain an overall factor
$e^{-d_{m,v}(C,C)}$,
where
$d_{m,v}(C,C)$ is an integral diverging to $+\infty$ as $m\to0$, unless
$C(e)\equiv\sum_iq_ic_i(e)=0$. In the limit, the factor converges to
zero unless $C(e)=0$:
\bea{lim} e^{-d_m(C,C)}\to \delta_{C,0}.\eea
This factor entails that states with different ``charge functions'' $C$ are
mutually orthogonal, and produces an uncountable number of
superselection rules. The physical
meaning of the charge functions is that of ``photon clouds'' attached to charged
particles \cite{MRS3}. States created by the exponential field acting on the vacuum can
formally be regarded as coherent photon states lying outside the
vacuum Fock space, and these coherent
states belong to inequivalent representations of the Maxwell field
whenever their photon clouds (i.e., their smearing functions $c(e)$) differ.

In \cite{MRS3}, a ``dressed Dirac field''
\bea{dd}\psi_{q,c}(x)=\wickv{e^{iq\phi(x,c)}}\cdot \psi_0(x)\eea
is introduced, where $q$ is the unit of electric charge. This field arises by subjecting
the free Dirac field to the ``trivial'' interaction density
\bea{dressd} q\cdot\pa_\mu\phi(x,c)j^\mu(x).\eea
Being a total derivative, \eref{dressd} does not contribute to the total action, and 
gives rise to a trivial scattering matrix. The non-perturbative
construction of the dressed Dirac field is meant as a  first step
towards the full perturbative QED, by splitting into two parts a QED interaction
density that can be defined on a positive-definite subspace of the
indefinite Fock space (i.e., a Krein space).
Although the interaction density \eref{dressd}
is ``trivial'', it drastically changes the algebraic structure
of the charged field. The dressed Dirac field is string-localized
(see \sref{s:pre}) and falls outside the
regime of, say, Wightman quantum field theory. Unlike the free Dirac field, it
creates states enjoying infrared features of QED that cannot be
attained in the usual local approach to QED (including the quantum
Gauss Law, the photon cloud superselection
structure, and the associated breakdown of Lorentz invariance). But
the dressed Dirac field does not interpolate between 
different scattering states. A nontrivial S-matrix is only produced
when also the ``true''
interaction $q\cdot A^K_\mu j^\mu(x)$ is turned on.

In some disguise, the exponentiated escort field has previously made
its appearance as a ``dressing factor'' in the Faddeev-Kulish
prescription \cite{FK} to prevent the formal vanishing of the LSZ
limit, and hence of the scattering matrix 
of QED \cite{Wein,Chu}. But  
(apart from further differences in the detailed formula), the dressing
factor in \cite{FK} is not
part of the charged field but rather of the states in which the
S-matrix has to be evaluated.  

Scattering theory in each of its formulations exploits the asymptotic large-time behaviour
of correlation
functions. Thus, a future modification of scattering theory adapted to
theories with infraparticles will need detailed information about 
correlation functions of the infrafield. In the case of the dressed
Dirac field, these involve correlation functions of the vertex
operators. This is one of our motivations to study the latter.

The new approach to QED itself is not the topic of  this paper,
except for the short remarks in \sref{s:pre}. For more, we refer to \cite{MRS3}.
Our topic are correlation functions of vertex operators and their
analytic properties. In particular, the coefficients $q$ in \eref{wickv} may be regarded as free
parameters, unrelated to the electric charge. The results also bear on
technical aspects of the very setup of string-localized quantum field
theory \cite{MSY}, e.g., how much smearing of the string directions is necessary.  

\section{Preliminaries}
\label{s:pre}

The basic idea of the new approach to QED \cite{MSY,S1,MRS2} is to use ``string-localized potentials'' $A_\mu(x,e)$
for the free Maxwell field: 
\bea{AeF} F_{\mu\nu}(x)=\pa_\mu A_\nu(x,e)-\pa_\nu A_\mu(x,e).\eea
They depend on a spacelike four-direction $e$, and enjoy the
axiality property
\bea{ax} e^\mu A_\mu(x,e)=0.\eea
It is suggestive to think of them as ``axial gauge'' potentials; but
the choice of $e$ is not a gauge-fixing condition; rather the potentials $A_\mu(x,e)$
for all $e$ coexist on the same Hilbert space where $F_{\mu\nu}$ is defined. More precisely, $A_\mu(x,e)$ is a
distribution also in the variable $e$ and requires a smearing with suitable functions
$c(e)$. The latter are required to have total weight $\int d\sigma(e)\,
c(e)=1$, so that $A_\mu(x,c)$ are still potentials for $F_{\mu\nu}$.

The terminology ``string-localized'' refers to the fact
that they can be defined as
integrals (well defined in the distributional sense) over the field along a ``string'' of direction $e$:
\bea{Ae}A_\mu(x,e):= \int_0^\infty ds\, F_{\mu\nu}(x+se)e^\nu.\eea
String-localization is an algebraic property: $A(x,e)$ commutes with $F(y)$ provided $y$
is spacelike separated from the string $x+\RR_+e$.
The definition \eref{Ae} ensures both \eref{AeF} and \eref{ax}. Thus,
``axiality'' is a consequence of localization along a string.

The Hilbert space for the free Maxwell field can be directly obtained from
the massless unitary Wigner representations of helicities $\pm1$ \cite{Wein}
without the detour through a potential, and can thus be seen as a
primary entity. But in perturbative QED one usually starts from a local
potential $A^K_\mu(x)$, say
in the Feynman gauge, that is defined on an indefinite Krein space, and then defines
\bea{AKF} F^K_{\mu\nu}(x)=\pa_\mu A^K_\nu(x)-\pa_\nu A^K_\mu(x).\eea
When the Krein space is reduced to the physical Hilbert space by the
Gupta-Bleuler (or BRST) prescription, the Maxwell field becomes equivalent to
the one on the Fock space over the Wigner representation; while the
potential $A^K_\mu$ ceases to exist. In contrast, the string-localized  potentials
$A_\mu(x,e)$ (being functionals of the Maxwell field) exist on both the Krein space and on the physical
Hilbert space.

Because on the Krein space both \eref{AeF} and \eref{AKF} hold, the
two potentials differ by an operator-valued gauge transformation
\bea{AeAK}
A^K_\mu(x,e) = A^K_\mu(x) + \pa_\mu \phi(x,e).
\eea
The quantity $\phi(x,e)$, baptized ``escort field'' in
\cite{S2}, turns out to be given by \eref{phiAK}.
Thus, it is string-localized and -- as we shall see -- infrared
divergent. But its derivative is well-defined as a string-integral
over $\pa A^K$ that decays fast enough as $s\to\infty$.

There are now several new options \cite{MRS3} to construct QED without the need to
work in Krein spaces. The first is to replace the usual interaction
density $q\, A^Kj$, defined on a Krein space, by $q\, A(e)j$, defined
on the Wigner Hilbert space (where the string-dependence is a total derivative and
should not affect the resulting theory \cite{S2}). This option requires
string-localized propagators of the potentials $A(e)$, that are the
topic of \sref{s:propa}.

The ``hybrid'' option 
indicated in the introduction is to split $q\, A(e)j$ into $q\,
\pa\phi(e)j$ (which is a total derivative) and $q\, A^Kj$, and study
the theory with the ``trivial interaction density'' 
$q\,\pa\phi(e)j$ first. This model can actually be constructed
non-perturbatively, leading to a rigorous and IR-finite definition of
the vertex operators \eref{wickv} and the dressed Dirac field
\eref{dd}. The analytic properties of vertex operators, in
particular, the space-time structure of their correlation functions,
is the topic of the main part of this paper, \sref{s:pre} and \sref{s:2pt}.

The second step, the perturbation of the dressed Dirac field with the
QED interaction density $q\, A^Kj$ (and the reasons why this does not
re-introduce Hilbert space non-positivity) 
requires a rather big effort and is addressed in \cite{MRS3}.

Let us begin with an inventory of the basic quantities that are needed
in the various approaches.

The two-point function $W_0$ and the Feynman propagator $G^F_0$ of the
massless scalar Klein-Gordon field
coincide (up to a factor of $i$) as a function of $x=x_1-x_2$ outside
the singular support, which is the null-cone $(x_1-x_2)^2=0$. 
As distributions, they are given in position space as boundary values
of analytic functions
\bea{W0}W_0(x)
=\lim_{\eps\downarrow0}\frac 1{(2\pi)^{2}}\frac
{-1}{(x^0-i\eps)^2-\vec x^2} &\equiv& \lim_{\eps\downarrow0}\frac
1{(2\pi)^{2}}\frac {-1}{x^2-i\eps x^0}\equiv \frac{-1}{(2\pi)^{2}}\cdot 
\frac {1}{(x^2)_-}, \\ \label{G0F} G_0^F(x) 
&=& \lim_{\eps\downarrow0}\frac 1{(2\pi)^{2}}\frac{-i}{x^2-i\eps},\eea
respectively. The commutator function (Pauli-Jordan function) can be written as
\bea{C_0} C_0(x):= i(W_0(x)-W_0(-x)) =
\frac{1}{2\pi}\,\sign(x^0)\delta(x^2).\eea
For the indefinite Feynman gauge vector potential, one has
\bea{fns} \erw{A^K_\mu(x)A^K_\nu(x')}=-\eta_{\mu\nu} W_0(x-x'),\\
i\erw{TA^K_\mu(x)A^K_\nu(x')}=-\eta_{\mu\nu} G_0^F(x-x').\eea
By \eref{phiAK}, the ensuing two-point function and Feynman propagator
of the escort field are given as double integrals
\bea{doub}-(ee')\int_0^\infty ds'\int_0^\infty ds \,F(x+se-s'e') \equiv -(ee')(I_{-e'}I_{e}F)(x), \qquad
(F=W_0\,\,\hbox{resp.}\,\,G_0^F).\eea
The notation $I_e$ stands for the string-integration as in
\eref{phiAK} or \eref{Ae}.  
The operations $I_e$ commute among each
other and with derivatives as long as all integrals exist as distributions, and it holds
\bea{edI} e^\mu\pa_\mu (I_ef)(x)= -f(x).\eea
If $f$ is a function or a distribution, $I_ef$ can only be
defined if $f$ has sufficiently rapid decay. Because $W_0$ and $G^F_0$ fall off in configuration space like $1/x^2$, their first
string-integrations are finite, whereas the second string-integration
diverges logarithmically and has to be regularized. 

The well-definedness as a distribution is a more
subtle issue than the convergence of an integral. In Fourier space,
the string-integration is a multiplication with another distribution:
\bea{Iekx}I_e e^{-ikx} \equiv \ioi ds\, e^{-ik(x+se)} =
\lim_{\eps\downarrow0}\frac {-i}{(ke)-i\eps}\cdot e^{-ikx}.\eea
The existence of a product of distributions has to be analyzed by
microlocal methods, such as Hörmander's criterion for the wave front
sets. But the latter is only a sufficient condition, and by
cancellations of singularities the product may be better behaved than
the wave front sets may tell. Since in this work we are
interested mainly in the behaviour in position space, we 
refer to \cite{G} where the existence of certain relevant
distributions has been established in Fourier space.

We just notice here that the Fourier transforms of $W_0$
and $G^F_0$ scale like $k^{-2}$. The IR divergence arises because
with two additional denominators as in 
\eref{Iekx}, the Fourier integrals would diverge logarithmically at $k=0$. 

Because $\frac {1}{(y-i\eps)^2}$ (where $y=(ke)$) is well-defined, $(I_eI_ef)(x)$ is
well-defined as a distribution in $x$, provided the decay of $f$ is fast
enough. But because $\frac 1{(y-i\eps)(y+i\eps)}$ is ill-defined,  $(I_{-e}I_ef)(x)$ is
always ill-defined. In position space, this can be easily understood
because the integrand in \eref{doub} depends only on $s-s'$. In particular, $\erw {A(x,e)A(x',e')}$ is ill-defined
at $e=e'$. Not least for this reason, one should consider the
string-integrals also as distributions in $e$ and $e'$. We shall see
in \sref{s:2pt-two} that the singularity at $e=e'$ is integrable in $e$ and $e'$ w.r.t.\ the invariant measure of $S^2$, so that smeared expressions like  $\erw {A(x,c)A(x',c')}$ are
well-defined even when the supports of $c(e)$ and $c'(e')$ overlap.

Because $W_0$ and $G_0^F$ are homogeneous distributions in $x$ of
degree $-2$, $I_eW_0$
and $I_eG_0^F$ are homogeneous both in $e$ and in $x$ of degree
$-1$. We shall (most of the time) restrict $e$ to the open set of spacelike vectors, because these are the
directions needed in the intended applications \cite{MRS3}. Because of
homogeneity in $e$, we may as well restrict $e$
to the unit spacelike hyperboloid
$H_1=\{e\in \RR^4:e^2=-1\}$ \cite{G}; but it will be advantageous to display factors
``$e^2$'' explicitly, so as to maintain homogeneity in $e$. Yet, smearing in $e$ is always understood with normalized $e$.

The quantities of interest in various applications \cite{MRS3} are:
\begin{enumerate}\itemsep-1mm
\item Two-point functions of $A_\mu(x,e)= (I_eF_{\mu\nu})(x)e^\nu$ or 
  $A^K_\mu(x,e)=A_\mu^K(x)+\partial_\mu\phi(x,e)$. (These distributions
  are identical in the Wigner space
  version and the Krein space version.)
\item   Mixed two-point functions between $A_\mu^K(x)$ and $\phi(x,e)$.
  \item IR-regularized two-point functions of $\phi(x,e)$ and their exponentials.
\item Propagators (= time-ordered two-point functions) of
  $A_\mu(x,e)$.
  \end{enumerate}
  Propagators of $\phi$ with itself would require
    time-ordering {\em and} IR regularization, which would be very delicate
    to implement simultaneously. Fortunately, such objects do not occur \cite{MRS3}. 

  (i) define a string-localized  quantum field theory. Apart from a local
  contribution, they involve contributions with one or two
  string-integrations over derivatives of   \eref{W0}. Thanks to the
  derivatives, these string-integrations are IR-finite and do not need a
  regularization. They are far simpler than (iii) without derivatives.

  (ii) and (iii) arise in the ``hybrid'' approach of \cite{MRS3},
  where the escort field without derivative appears  in the exponent
  of a regularized normal-ordered Weyl operator (vertex operator). 
  Their computation and analysis is our first main topic.

    (iv) are needed in perturbation theory when a current is coupled to
  a string-localized potential. They involve one or two string-integrations over
  derivatives of \eref{G0F}. Again, thanks to the derivatives, an IR
  regularization is not needed. Their computation and analysis is the
  second main topic of this work.

    Thus, we shall study the IR-finite expressions
    \bea{IRfin}(I_eW_0)(x), 
 \quad (I_{e_2}I_{e_1}\pa W_0)(x),  \quad (I_{e_2}I_{e_1}\pa G_0^F)(x)   \eea
 from which the other IR-finite quantities of interest arise via 
\bea{IeW} (I_e\pa W_0)(x)= \pa (I_eW_0)(x),\qquad (I_{e_2}I_{e_1}\pa \pa
W_0)(x)=\pa (I_{e_2}I_{e_1}\pa W_0)(x);\eea
and the IR-regularized expression
\bea{IR-reg} (I_{e_2}I_{e_1}W_0)_v(x).
\eea

\section{Two-point function}
\label{s:2pt}

\subsection{One string-integration}
The distribution $W_0(x)$ is defined as the boundary value of the
analytic function $-\frac1{(2\pi)^2}\frac1{z^2}$ in the complex forward tube
$\IM z^0<0$. We may thus write $z=x-i\eps u$ where $x$ is real and $u$
a forward timelike (unit) vector.  By Lorentz invariance, the distributional  limit $\eps\downarrow 0$ is
independent of $u$. So, because $e$ is spacelike, one may choose $u$
perpendicular to $e$. Let in this section $F(x)= \frac 1{(x-i\eps u)^2}$, where the distributional limit $\eps\downarrow 0$ is always
understood.

\label{s:2pt-one}
The string-integral over $F(x)$ can be written as
\bea{I1s}  f(x,e):=-(I_eF)(x)\equiv -\ioi \frac{ds}{(x+se-i\eps
  u)^2} = -\ioi \frac{ds}{(x-i\eps u)^2+2s(xe) + s^2e^2}.\eea
The point is that the complex denominator cannot vanish for real $s$,
and $i\eps$ appears only in the parameter $(x-i\eps u)^2$. The 
elementary integration gives
\begin{lemma}\label{l:fxe}
  \bea{fxe}
(2\pi)^2 (I_eW_0)(x)=f(x,e) =
\frac{\frac12\log\frac{-(xe)+i\sqrt{\det_{x,e}}}{-(xe)-i\sqrt{\det_{x,e}}}}{i\sqrt{\det_{x,e}}}=\frac{\frac12\log\frac{-(xe)+\sqrt{-\det_{x,e}}}{-(xe)-\sqrt{-\det_{x,e}}}}{\sqrt{-\det_{x,e}}},\eea
where 
\bea{detxe}\det{}_{x,e}:= (x-i\eps u)^2e^2-(xe)^2\eea
is the Gram determinant, with the imaginary shift of $x$ being implicitly understood.
\end{lemma}
The logarithm of the quotient is understood as the difference of two
logarithms with their branch cuts along
$\RR_-$.\footnote{\label{fn:log}In particular, negative factors must not be
  cancelled under the logarithm: this would produce errors of $2\pi i$!}
In particular, \eref{fxe} does {\em not} depend on the choice of the branch of the
square root (because numerator and denominator would simultaneously
switch sign). As a function on $\RR^4\times H_1\times H_1$ (i.e., putting $\eps=0$), \eref{fxe} is
ill-defined only when $x^2=0$ or $\det_{x,e}=0$. The $i\eps$-prescription
in $\det_{x,e}$ defines \eref{fxe} as a distribution.

When $x$ and $e$ lie in a common spacelike plane, one may without loss
of generality assume $x^0=0$ and $e^0=0$. Then
\bea{fxe-spl} \sqrt{x^2e^2}\cdot f(x,e) =
\frac{\alpha}{\sin\alpha},\eea
where $\alpha=\angle(\vec x,\vec e)\in [0,\pi)$. 
It has a singularity at $\alpha=\pi$ reflecting the fact that the
string $x+\RR_+e$ passes through the origin. Yet, upon smearing in
$\vec e$, this singularity is integrable w.r.t.\ the invariant measure
of $S^2$ and does not need an $i\eps$
prescription to define it as a distribution in $\vec e$.

More generally, the defining integral \eref{I1s} may be singular
whenever $(x+se)^2$ can become zero for $s\geq0$, i.e., geometrically,
when the string $x+\RR_+e$ hits the null-cone. This happens
necessarily if $x$ is timelike or lightlike (the string starts
inside or on the null-cone). If $x$ is spacelike, the string may touch the null-cone  or pierce it
twice. However, inspection of \eref{fxe} shows that with the
$i\eps$-prescription, the singular support of $f(x,e)$ is at $x^2=0$
(the string starts on the null-cone) and at $\det_{x,e}=0$,
$(xe)\geq0$ (the string touches the null-cone). 

Somewhat unexpected from its unsymmetric definition $f(x,e)=-(I_eF)(x)$,
this function (where it is regular) is symmetric in $x\leftrightarrow e$. The
symmetry can be understood by a change of integration variables
$s\to\frac1{s}$ in \eref{I1s}. 

\subsection{Two string-integrations}
\label{s:2pt-two}

The two-point function of the escort field is the twofold string-integral over $W_0(x)$, multiplied by the factor $-(ee')$. The
presence of this factor jeopardizes the positivity of the inner product defined by the two-point
function, which is essential on the way to the superselection
structure via \eref{exp-d}, outlined in the introduction. To secure
positivity, one has to impose that $e$ and $e'$ are smeared within a
spacelike surface \cite{MRS3} perpendicular to a timelike unit
vector $u$. Without loss of generality, we may pick
\bea{u0} u=u_0:=\bpm 1\\ \vec 0\epm \qquad \Rightarrow \qquad e=\bpm 0\\
\vec e\epm, \quad e'=\bpm 0\\
\vec e\,'\epm.\eea
Thus, smearing functions $c(e)=c(\vec e)$ are elements of $C^\infty(S^2)$.
Vectors $y$ with $y^0=0$ will be called ``purely spatial''.

Because the two-point function $W_0(x)$ is homogeneous of degree $-2$,
$(I_eW_0)(x)$ is homogeneous of degree $-1$, and the second string-integration would 
diverge logarithmically. In momentum space,
\bea{IIWmom} (I_{-e'}I_e W_0)(x)\stackrel?=\lim_{\eps\downarrow0}\lim_{\eps'\downarrow0}\int d\mu_0(k)
\frac{e^{-ikx}}{((ke)-i\eps)((ke')+i\eps')}\eea
diverges at $k=0$. We therefore have to regularize it in the
infrared. The regularization extends the momentum space distribution
to $k=0$. This is done by replacing $e^{-ikx}$ by $e^{-ikx}-v(k)$ where $v(k)$
is any smooth test function with $v(0)=1$. Thus, we define
\bea{IvWdef}(I_{-e'}I_e W_0)_v(x):=\lim_{\eps\downarrow0}\lim_{\eps'\downarrow0}\int d\mu_0(k)
\frac{e^{-ikx}-v(k)}{((ke)-i\eps)((ke')+i\eps')}.\eea
 Because of the symmetry $e\lra -e'$, we shall in the
 sequel write $e=e_1$ and $-e'=e_2$, so that
 $(I_{e_2}I_{e_1}W_0)_v(x)$ is symmetric in $e_1\lra e_2$.

We want to gain insight into the distribution $(I_{e_2}I_{e_1} W_0)_v(x)$
 in position space, leaving the regulator function $v$
 unspecified. It is
 clearly not possible to compute the integral \eref{IvWdef} when
 $v(k)$ is not specified.
The strategy is therefore to compute instead the cut-off integral
 \bea{IaIdef}(I^a_{e_2}I_{e_1} W_0)(x)\equiv \int_0^a ds_2\,
 (I_{e_1}W_0)(x+s_2e_2) = \frac1{(2\pi)^2} \int_0^a ds\, f(x+se_2,e_1)
 \eea -- which can be done analytically --  and use that
\bea{IvIa}\pa_\mu (I_{e_2}I_{e_1} W_0)_v(x) = (I_{e_2}I_{e_1} \pa_\mu W_0)(x) = \lim_{a\to\infty} (I^a_{e_2}I_{e_1} \pa_\mu
W_0)(x) = \lim_{a\to\infty} \pa_\mu (I^a_{e_2}I_{e_1} 
W_0)(x),\eea
where the first equality follows from the definition \eref{IvWdef},
the second holds because $\pa_\mu W_0$ decays sufficiently fast to
make the integral converge in $a$, and the last is obvious from the
definition of the integral operations. Thus, the difference is
independent of $x$, and the result for $(I_{e_2}I_{e_1}W_0)_v(x)$ is obtained by 
replacing the cut-off dependent but $x$-independent term by another
(unknown) $x$-independent term. Specifically, we will show in the
remainder of this section 
\bea{IaI}(I^a_{e_2}I_{e_1} 
W_0)(x) = \frac{1}{(2\pi)^2} \Big[\frac12 f(e_1,e_2)\cdot
\log\big(\frac{4(ae_2)^2}{(x-i\eps u)^2}\big) +
\frac{H(x;e_1,e_2)}{(e_1e_2)}\Big] + O(\frac1a),\eea
where $f(e_1,e_2)=f(e_2,e_1)$ is the same distribution as in \eref{I1s}, and
the distribution $H$ is symmetric in $e_1\lra e_2$ and homogeneous of
degree $0$ separately in $x$, $e_1$, and $e_2$. We conclude 
\bea{IvW}(I_{e_2}I_{e_1}W_0)_v(x) = \frac{1}{(2\pi)^2} \Big[-\frac12f(e_1,e_2)\cdot \log(-\mu^2_v\cdot(x-i\eps u)^2) +
\frac{H(x;e_1,e_2)}{(e_1e_2)}\Big],\eea
where $\mu_v$ carries the
dependence on the regulator function $v$ and may depend on $e_1$ and $e_2$.

\newpage

To prepare the computation of $H(x;e_1,e_2)$ in \eref{IaI} and
\eref{IvW}, we need some definitions. We shall denote by
\bea{Det} \det{}_{y_1,y_2,y_3}:=y_1^2y_2^2y_3^2 - y_1^2(y_2y_3)^2 - y_2^2(y_1y_3)^2 -
y_3^2(y_1y_2)^2 + 2 (y_1y_2)(y_1y_3)(y_2y_3)
\eea
the Gram determinant of vectors $y_1$, $y_2$, $y_3$. For
$i,j,k\in\{1,2,3\}$ pairwise distinct, the cofactors of $y_i^2$ are
the $2\times2$ Gram determinants $\det_{y_j,y_k}$, and we shall denote the
cofactors of $(y_iy_j)$ by
\bea{Lambda} \Lambda_k = (y_iy_k)(y_jy_k)-y_k^2(y_iy_j).\eea
We shall need a few trivial facts, proven by elementary computation.
\begin{lemma}\label{l:Ginv} It holds
  \bea{L}\partial_{y_i}\det{}_{y_1,y_2,y_3} = 2 \bpm \det_{y_2,y_3}&\Lambda_3&\Lambda_2\\
  \Lambda_3&\det_{y_1,y_3}&\Lambda_1\\ \Lambda_2 &\Lambda_1&
  \det_{y_1,y_2}\epm \bpm y_1\\ y_2\\ y_3\epm.\eea
 If $G$ is the Gram matrix and $L$ the matrix in \eref{L}, then
 $GL=LG=\det_{y_1,y_2,y_3}$, i.e., if\ $\det_{y_1,y_2,y_3}\neq0$, then
 $\det_{y_1,y_2,y_3}\inv L = G\inv$.
  \end{lemma}
\begin{lemma}\label{l:MM} For $i,j,k\in\{1,2,3\}$ pairwise distinct, it holds
  \bea{DMM}y_i^2\det{}_{y_i,y_j,y_k}=\det{}_{y_i,y_k}\det{}_{y_i,y_j} - \Lambda_i^2.\eea
  \end{lemma}
Notice  that with Lorentzian metric the vanishing of a Gram
determinant does not require the linear dependence of the
vectors, see, however \lref{l:det30} and \lref{l:det20}. 

The inverse of the Gram determinant $\det{}_{x,e_1,e_2}$ will play a major
role. It is 
understood  as the distributional boundary value from the forward
tube $x-i\eps u$. Because $(e_iu)=0$, this simply means that $x^2$ is
understood as $(x-i\eps u)^2$ while all other scalar products are
real. For properties of Gram determinants in Lorentzian metric, see \sref{s:propa}.

It is convenient to define
$\gamma=\angle(\vec e_1,\vec
e_2)$, so that $\sqrt{e_1^2e_2^2}\,\cos\gamma = -(e_1e_2)$ and
$\det_{e_1,e_2}=e_1^2e_2^2\sin^2\gamma$. One trivially has
\begin{lemma}\label{l:fee}
The distribution $f(x,e_2)$ in \lref{l:fxe} with $x$ substituted by
$e_1$ equals
\bea{fee} \sqrt{e_1^2e_2^2}\cdot f(e_1,e_2) = \frac\gamma{\sin\gamma}.\eea
The singularity at $\gamma=\pi$ is integrable w.r.t.\ the invariant
measure on $S^2\times S^2$. 
\end{lemma}
\begin{remk} \label{r:smear} The singularity is {\em not} integrable along 
one-dimensional submanifolds of $S^2$. Thus, strings must not be
further restricted than $e^2=-1$ (which is trivial by homogeneity) and $e^0=0$.
\end{remk}

Next, by using \lref{l:MM} with $y_1=e_1$, $y_2=e_2$, $y_3=x$, and
$i=1$ and $i=2$,
respectively, one can define the homogeneous functions 
$\zeta_1(x,e_1,e_2)$ and $\zeta_2(x,e_1,e_2)$ by
\bea{zeta} \pm e^{\pm \zeta_1} = \frac{\Lambda_1\pm
  \sqrt{\det_{e_1,e_2}\det_{x,e_1}}}{\sqrt{e_1^2\det_{x,e_1,e_2}}}, \qquad
\pm e^{\pm \zeta_2} = \frac{\Lambda_2\pm
  \sqrt{\det_{e_1,e_2}\det_{x,e_2}}}{\sqrt{e_2^2\det_{x,e_1,e_2}}}.
\eea
When $e_1$, $e_2$ and $x$ are purely spatial, the geometry is Euclidean. Then all
diagonal cofactors ($2\times 2$ Gram determinants) are $\geq0$ and
$e_1^2\det_{x,e_1,e_2}\geq0$. In this case, $\zeta_1$ and $\zeta_2$ are real.  The same is true when
$x^\perp$, the component of $x$ perpendicular to the plane spanned by
$e_1$ and $e_2$, is
spacelike, because such $x$ can be boosted to become purely spatial while
$e_1$ and $e_2$ are preserved.

We can now state the result.
\begin{prop}\label{p:H} Let $e_i$ ($i=1,2$) be purely spatial
  and linearly independent ($\det_{e_1,e_2}\neq0$). Denote
  by $D=\frac{\det_{x,e_1,e_2}}{(x-i\eps u)^2e_1^2e_2^2}$ the
  normalized Gram determinant. Then
  the distribution $H$ in \eref{IaI} and 
  \eref{IvW} is 
\bea{H} H(x;e_1,e_2) = -\frac{\cos\gamma}{2\sin\gamma}\left[\gamma\log\big(\frac{\sin^4\gamma}{D}\big)+
\pi(\zeta_1+\zeta_2)-\frac
i{2}\left\{\Li_2\Big(e^{i\gamma}e^{\zeta_1}e^{\zeta_2}\Big)\ba{c}+(e^{\zeta_1}\lra
  -e^{-\zeta_1})\\ +(e^{\zeta_2}\lra -e^{-\zeta_2})\\-(e^{i\gamma}\lra
  e^{-i\gamma})\ea\right\}\right]. \quad
\eea
The limit $\gamma\to0$ ($e_1$ and $e_2$ parallel) is regular, while the limit $\gamma\to\pi$ ($e_1$ and $e_2$
antiparallel) is singular, but integrable w.r.t.\ the invariant
measure on $S^2\times S^2$. 
\end{prop}
Via \eref{IvW}, this formula determines the regularized double string-integrated
distribution $(I_{e_2}I_{e_1}W_0)_v(x)$ on $\RR^4\times S^2\times
S^2$, up to the
unknown additive dependence on the regulator function $v$ via
$\mu_v(e_1,e_2)$. Its relevance for the intended applications to QED will
be discussed at the end of \sref{s:vertex}.

{\em Sketch of proof:} For \eref{IaI} we need to compute the integral over
$f(x+se_2,e_1)$ in \eref{IaIdef}.
We begin with $x^\perp$ (the component of $x$ perpendicular to $e_1$ and $e_2$)
spacelike. Then there is a boost preserving $e_1$ and
$e_2$, such that $(u\Lambda x)=0$. Thus, we may without loss of
generality assume that also $x$ is purely spatial, and $\zeta_1$ and
$\zeta_2$ are real.

The clue to compute \eref{IaIdef} analytically is the change of integration variable
\bea{C}C(s) = e^{\arcsinh(s\Gamma_1+\Gamma_2)}= s\Gamma_1+\Gamma_2 + \sqrt{(s\Gamma_1+\Gamma_2)^2+1},\eea
where 
\bea{Gamma}\Gamma_1:=\frac{\det_{e_1,e_2}}{\sqrt{e_1^2\det_{xe_1e_2}}},\qquad
\Gamma_2:=-\frac{\Lambda_1}{\sqrt{e_1^2\det_{xe_1e_2}}}\eea
with $\Lambda_i$ defined as in \eref{Lambda} for $y_1=e_1$, $y_2=e_2$, $y_3=x$.
Then $C(0)=e^{-\zeta_1}$ and $C(a)= \log(2a\Gamma_1) +
O(\frac1a)$ for large $a$, and one finds 
\bea{fxse} f(x+se_2,e_1) =
\frac{\Gamma_1}{\sqrt{\det_{e_1,e_2}}}\frac{\log\frac{\frac{e^{-\zeta_2}}{2\Gamma_1}(1-e^{i\gamma}e^{\zeta_2}C(s))(1+e^{-i\gamma}e^{\zeta_2}C(s)\inv)}{\frac{e^{-\zeta_2}}{2\Gamma_1}(1-e^{-i\gamma}e^{\zeta_2}C(s))(1+e^{i\gamma}e^{\zeta_2}C(s)\inv)}}{i(C(s)+C(s)\inv)}.\eea
Because $\zeta_i$ are real and $\Gamma_1$ is positive, one can 
cancel the factor
$\frac{e^{-\zeta_2}}{2\Gamma_1}$ in the difference of logarithms without
risk of changing the branches, see footnote \ref{fn:log}. The integral is solved by
\bea{prim}
2\Gamma_1\frac{\log(1-e^{i\gamma}e^{\zeta_2}C(s))(1+e^{-i\gamma}e^{\zeta_2}C(s)\inv)}{C(s)+C(s)\inv}
= \frac d{ds}\Big[\Li_2(-e^{-i\gamma}e^{\zeta_2}C(s)\inv)-\Li_2(e^{i\gamma}e^{\zeta_2}C(s))\Big],\qquad\eea
and likewise for $\gamma\to -\gamma$. By working out the values at
$s=0$ and $s=a$, and writing the result symmetrically in $e_1\lra
e_2$, one arrives at \eref{H}.

The
behaviour at the branch cuts of the dilogarithm function, determined by the imaginary
part of $x^0$,  can be worked out explicitly in both limits
$\gamma\to0$ and $\gamma\to\pi$. The singularity at $\gamma=\pi$ is like
$\frac{O(\log(\pi-\gamma))}{\pi-\gamma}$ and hence integrable.

When $x$ is not spacelike, the variables and functions are defined by
analytic continuation in $x-i\eps u$ in the forward tube, and the distribution as the
boundary value $\eps\downarrow0$. The claims follow by analytic continuation
in $x^0-i\eps$ in the forward
tube.\qed

The regular behaviour at $\gamma=0$ ($e_1=e_2$) is expected because
$(I_e^2F)(x)=\ioi s\, ds \, F(x+se)$ is well-defined when regularized as
in \eref{IvWdef}. The singular behaviour at $\gamma=\pi$ ($e_1=-e_2$)
is expected because $I_{-e}I_e$ is never defined.

\begin{coro}\label{c:phiphi}
Let $\wt f(e,e'):=f(e,-e')$ and $\wt H(x-x';e,e'):= H(x-x';e,-e')$, which are symmetric under $e\lra-e'$. The regularized two-point function of the escort field 
\bea{phiphi}
 (2\pi)^2 \erw{\phi(x,e)\phi(x',e')}_v = \frac{(ee')}2 \wt f(e,e')
 \log\big(-\mu_v^2\cdot(x-x'-i\eps u)^2\big)+\wt H(x-x';e,e')\quad\eea
 is a distribution on $\RR^4\times S^2\times S^2\setminus\{(x,e,e'):e=e'\}$.
   \end{coro}

When smeared with the constant function $c_0(e_i)=\frac1{4\pi}$,
\eref{phiphi} simplifies drastically. The averaging with $c_0$ can be
done already in the momentum space representation \eref{IvWdef}: 
\begin{lemma}\label{l:T0} (see \cite{MRS3}) For
purely spatial $e$, $u^0$ the standard timelike unit vector as above, and $k^2=0$ one has
  \bea{T0}\int_{S^2} d\sigma(\vec e)\, c_0(\vec e) \,\frac{e}{(ek)_\pm} =
\frac{u_0}{(u_0k)} - \frac{k}{(u_0k)^2}.\eea
\end{lemma}
Because $(\frac{u_0}{(u_0k)} - \frac{k}{(u_0k)^2})^2 = -\frac
1{(u_0k)^2}$, which is the denominator for two string-integrations in
the direction $u_0$, one obtains the desired result by computing
$(I_{u_0}I_{u_0}W_0)_v$ along the same lines as
$(I_{e_2}I_{e_1}W_0)_v$ before. The result
is 
\begin{lemma}\label{l:phi0phi0} For $c_0(\vec e)=\frac 1{4\pi}$, one has
\bea{phi0phi0} 
 (2\pi)^2 \erw{\phi(x,c_0)\phi(x',c_0)}_v& =& -\frac12
 \log\big(-\wt\mu_v^2\cdot((x-x')^2)_-\big)+\wt H(x-x';c_0,c_0), \\
 \hbox{with}\qquad \wt H(x;c_0,c_0) &=&\frac {x^0}{2r}
 \log\frac{(x^0-i\eps)+r}{(x^0-i\eps)-r} \qquad (r=\vv x).\eea
\end{lemma}
{\em Sketch of proof:} By a direct computation of the cut-off integral \eref{IaI},
which for two parallel strings becomes an elementary integral. The
claim for the regularized integral follows by the argument preceding
\eref{IaI}, which leaves the constant $\wt\mu_v$ unspecified.\qed

\subsection{Vertex operator correlations and commutation relations}
\label{s:vertex}
We briefly sketch the definition of operators
$\wickv{e^{i\phi(g\otimes c)}}$ (smeared in both $x$ and $e$) through a massless
limit \cite{MRS3}:
\bea{eiphiv} \wickv{e^{i\phi(g\otimes c)}} :=\lim_{m\to0} e^{-\frac{\wh g(0)^2}2 d_{m,v}(c,c)}\cdot
\wick{e^{i\phi_m(g\otimes c)}} \equiv \frac{\wick{e^{i\phi(g\otimes c)}}}{e^{-w_v(g\otimes c,g\otimes c)}},
\eea
where $\phi_m$ is the IR-regular massive escort field whose two-point
function diverges in the limit $m\to0$, and
\bea{dee} d_{m,v}(e,e')  = -(ee')\int \frac{d\mu_m(k)
  \,v(k)}{((ke)-i\eps)((ke')+i\eps)},\eea
smeared over $e$ and $e'$ is the divergent part. The second writing in
\eref{eiphiv} refers to a normal-ordering prescription w.r.t.\ the
non-positive two-point function $w_v$ given by \eref{phiphi}. However,
correlations of \eref{eiphiv} evaluated in the limit of massive vacuum
states define a positive functional. The correlations can be worked
out by the Weyl formula.  The crucial feature is that they contain the 
IR divergent parts in the combination
\bea{exp-d}e^{-\frac12 d_{m,v}(C,C)}\eea
where $C(e) = \sum_i \wh g_i(0)c_i(e)$. Because $d_{m,v}(C,C)$  diverges
to $+\infty$ unless $C=0$, one obtains in the limit a Kronecker delta
$\delta_{C,0}$. This defines an uncountable superselection rule: the
GNS Hilbert space splits into an uncountable direct sum of
subspaces $\mathcal {H}_C$ which carry inequivalent representations of the Weyl
subalgebra generated by $e^{i\phi(g\otimes c)}$ with $\wh g(0)=0$.

Therefore, correlation functions of $\wickv{e^{i\phi(g\otimes c)}}$ are
given by
\bea{wickcorr}\Erw{\wickv{e^{i\phi(g_1\otimes c_1)}}\dots
  \wickv{e^{i\phi(g_n\otimes c_n)}}}
= \delta_{C,0}\cdot \prod_{i<j} e^{-w_v(g_i\otimes c_i,g_j\otimes c_j)}.
\eea
When $w_v$ given by \eref{phiphi} is inserted, one may collect the 
contributions from $\log\mu_v^2(e,-e')$ in the exponent as 
\bea{sumlamb}\sum_{i<j} \frac{\wh g_i(0)\wh g_j(0)}{8\pi^2}\lambda_v(c_i,c_j) 
\eea
with the symmetric bilinear form
\bea{lambda}\lambda_v(c,c'):=\int d\sigma(\vec e)\,c(\vec e)\int d\sigma(\vec
e\,')c'(\vec e\,')\,\frac{\pi-\theta(\vec e,\vec
  e\,')}{\tan\theta(\vec e,\vec e\,')}\log\mu_v(e,-e'),\eea
where $\theta(\vec e,\vec e\,') = \angle(\vec e,\vec e\,')$. 
Because $\sum_i \wh g_i(0)c_i=0$, the exponential of the sum factorizes:
\bea{lambfact} \exp \sum_{i<j} \frac{\wh g_i(0)\wh
  g_j(0)}{8\pi^2}\lambda_v(c_i,c_j) = \prod_i e^{-\frac{\wh
    g_i(0)^2}{16\pi^2}\lambda_v(c_i,c_i)},\eea
and one may conveniently absorb the factors as $c$-dependent normalizations of the fields.

Vertex operators are defined by choosing $g(y)=q\delta_x(y)$, hence $\wh g(0)=q$:
\bea{V} V_{qc}(x):= N_v(c)\cdot
\wickv{e^{iq\phi(x,c)}}, \qquad (N_v(c)=e^{\frac{q^2}{16\pi^2}\lambda_v(c,c)})\eea
\begin{coro}\label{c:VVV} (see \cite{MRS3}) The correlation functions of vertex
  operators are 
\bea{VVV}
\Erw{V_{q_1c_1}(x_1)\dots V_{q_nc_n}(x_n)} = \delta_{\sum_i q_ic_i,0}\cdot
\prod_{i<j}
\Big(\frac{-1}{(x_i-x_j)^2_-}\Big)^{-\frac{q_iq_j}{8\pi^2}\erw{c_i,c_j}}
e^{-\frac{q_iq_j}{4\pi^2}\wt H(x_i-x_j;c_i,c_j)}, \eea
where
\bea{cc}\erw{c,c'}:= \int d\sigma(\vec e)\,c(\vec e)\int d\sigma(\vec
e\,')c'(\vec e\,')\,\frac{\pi-\theta(\vec e,\vec e\,')}{\tan\theta(\vec e,\vec e\,')}.\eea
\end{coro}
{\em Proof:} The statement follows by combining the definition
\eref{V} with \eref{wickcorr}, where $w_v$ is specified by \eref{phiphi}
  and \pref{p:H}. The formula \eref{cc} uses \eref{fee}.\qed

We conclude this section with some miscellaneous results about 
correlations of vertex operators, with only indications of proofs.

\subsubsection{Commutation relations}
\label{s:commrel}

\begin{prop}\label{p:any}
  The vertex operators satisfy anyonic commutation relations
  \bea{any}
V_{qc}(x)V_{q'c'}(x') = e^{iqq'\beta(x-x';c,c')}\cdot
V_{q'c'}(x')V_{qc}(x),\eea
where $\beta(x-x';c,c')$ arises by smearing with $c(e)$ and $c'(e')$
the escort commutator function 
\bea{beta}\beta(x-x';e,e') = i[\phi(x,e),\phi(x',e')] =- (ee')
(I_{-e'}I_e C_0)(x-x').\eea
 \end{prop}
{\em Proof:} The claim follows from the fact that vertex operators are
defined as limits of multiples of Weyl operators, \eref{eiphiv} and \eref{V}. The escort commutator function
does not suffer from the IR divergence because the Fourier transform of $C_0$
vanishes at $k=0$. \qed

\newpage

For $x^0=x'^0$, \eref{beta} vanishes because the equal-time commutator
vanishes and $e$, $e'$ are purely spatial. Otherwise it is a rather simple geometric quantity in terms of the
intersection of the
null-cone with the planar wedge $x-x'+\RR_+e-\RR_+e'$, which is a subset of a circle. It can be written
in a symmetric form, by writing $e_1=e$ and
$e_2=-e'$ as before, so that $\gamma=\angle(\vec e_1,\vec e_2)$:
\begin{lemma}\label{l:arc} (see \cite{R}) Denote by $A$ the total arc-length of the intersection of
the null-cone with the planar wedge $x+\RR_+e_1+\RR_+e_2$ of opening angle $\gamma$. Then
\bea{betaarc}\beta(x;e_1,e_2) = -\sign(x^0)\frac{A}{4\pi \tan\gamma} . \eea
\end{lemma}
{\em Proof:}
Perform the double
string-integral in \eref{beta} in Euclidean polar coordinates
$(r,\varphi)$ of the plane $x+\RR e_1+\RR e_2$. The
change of coordinates contributes the Jacobi determinant $\frac
1{\sin \gamma}$. Using \eref{C}, the relevant integral can be written
as
\bea{arc}\int r\,dr\, d\varphi \, \chi_W(r,\varphi) \delta(r^2-x^\perp{}^2) =
\frac 12 A, \eea
where $\chi_W$ is the characteristic function of the wedge, and
$x^\perp$ is the component of $x$ perpendicular to the
plane. Finally, $(e_1e_2)=-\cos\gamma$. Collecting all factors, yields
\eref{betaarc}. \qed

The
intersection of the wedge with the null-cone is empty if the two
strings are spacelike separated, in particular when
$x^\perp{}^2<0$. In this case $A=0$ and the
vertex operators commute.

\subsubsection{Spectrum}
\label{s:spec}

One would like to know the Fourier transform of the two-point function of
vertex operators because its
$c$-dependent energy-momentum distribution supported in
the interior of $V^+$ reveals the energy-momentum spectrum of the
  state created by the vertex operator \cite{MRS3}. It has to be added to the
mass-shell energy-momentum of the free Dirac particle. The sum is the
spectrum of the infraparticle state generated by the dressed Dirac field \eref{dd}. 

We do not know how to compute this Fourier transform for general $c$,
other than by working
out an exponential series of convolution products of \eref{phiphi}
(which is impractical). However, the case of the constant
smearing function $c(\vec e)=c_0=\frac 1{4\pi}$, for which the
two-point function drastically simplifies, see \lref{l:phi0phi0},
allows to quantify
the ensuing dissolution of the mass-shell in a special case.

 By \eref{phi0phi0}, the two-point function of the 
 vertex operator $V_{qc_0}$ is
\bea{V0V0}\Erw{V_{qc_0}^*(x_1)V_{qc_0}(x_2)} =
\Bigg[\frac
{\Big(\frac{x^0-r-i\eps}{x^0+r-i\eps}\Big)^{\frac{x^0}{
      r}}}{-( x^2 -i\eps x^0)}\Bigg]^{\frac{\alpha}{2\pi}} , \qquad
(x=x_1-x_2, \, r=\vv x).
\eea
Here, $\alpha:=\frac{q^2}{4\pi}$, which in the application to QED
\cite{MRS3} is the fine structure constant.

 \eref{V0V0} equals
  $(ix^0+\eps)^{-\frac\alpha\pi}$ multiplied with a power
  series in $\frac {r^2}{x_0^2}$. This structure allows to extract
  quantitative details of the Fourier transform, and
  hence of the rotationally invariant energy-momentum distribution
  $\rho(\omega,\vec k)$ in the state created by $V_{qc_0}$ \cite{R}. 
By putting $r=0$, one concludes that the distribution $\rho(\omega)=\int
  d^3k\,\rho(\omega,\vec k)$ of energies 
  decays like $\omega^{\frac\alpha\pi-1}$. By applying powers of the Laplacian 
  before putting $r=0$, one can compute averages of powers of $\vv k^2$ at fixed
  energy $\omega$. E.g., the average of the invariant masses 
  $\omega^2-\vv k^2$ at given energy $\omega$ is found to be 
  $\frac\alpha\pi\cdot\omega^2 + O(\alpha^2)$ with variance
  $\frac49\frac\alpha\pi\cdot\omega^4 + O(\alpha^2)$. These data are
  roughly compatible with an inverse-power-law distribution
  $\sim(\omega^2-\vv k^2)^{1-\frac\alpha\pi}$ near the mass-shell $\vv k = \omega^2$.

\subsubsection{Scattering}
\label{s:scatt}

Vertex operator correlation functions \eref{VVV} involve exponentials of the
distribution $\wt H$ (smeared in the strings), and the same is
true for correlations of the dressed Dirac
field \eref{dd}. Asymptotic properties of $\wt H$ at large times  will
become relevant in a future scattering theory. At this moment, a scattering
theory for infrafields like the dressed Dirac field has not yet been
formulated. The LSZ method fails because of the absence of a sharp mass-shell, and for the same reason, ``asymptotic creation operators''
needed in the Haag-Ruelle theory have not been found. Apart from the
lack of the mass-shell, a major 
obstacle is the product structure of vertex operator correlations that
is very different from that of free correlations. It probably entails that
``two-infraparticle states'' are not tensor products as in the Fock
space. 

A toy model without a mass gap would be a scattering theory for vertex
operators without the Dirac field. A candidate would be a version of Buchholz'
scattering theory for massless waves \cite{Bu1} that is successful in the two-dimensional
vertex operator model (no $\wt H$-terms) and yields an S-matrix that is
a complex phase \cite{DM}. If the same method
is naively applied in the case at hand, the $\wt H$-terms in the correlations
\eref{VVV} can produce a result of modulus $>1$, which jeopardizes the direct
interpretation of the relevant asymptotic limit as a scattering amplitude \cite{T}.
Nevertheless, the computation illustrates how features of $\wt H$ will play a
role in scattering theory.

The would-be scattering amplitude is a limit of four-point correlations
relative to two-point correlations, when the positions go
to infinity like $x\pm t\ell$, $x'\pm t\ell'$ in future and past lightlike
directions. Let  $\ell$, $\ell'$ be two non-parallel future directed lightlike
vectors. Consider the quantity
\bea{tlim} \lim_{t\to\infty} \frac{\Erw{V_c(f_{t\ell})^*V_{c'}(f'_{t\ell'})^*V_{c'}(f'_{-t\ell'})V_c(f_{-t\ell})}}{\Erw{V_c(f_{t\ell})^*V_c(f_{-t\ell})}\Erw{V_{c'}(f'_{t\ell'})^*V_{c'}(f'_{-t\ell'})}}
\eea
with smearing functions $f_{t\ell}(x)=f(x-t\ell)$ etc.\ shifted in
lightlike directions. To simplify matters, we absorb the charge factor
$q$ into the string smearing functions $c$ which therefore have
arbitrary total weight $q$.
\begin{lemma}\label{l:NoGo}
  Let $\ell$, $\ell'$ be future directed lightlike
  vectors, and $c,c'\in C^\infty(S^2)$. Then in the limit
  $t\to\infty$, \eref{tlim} converges to
  \bea{limit}e^{-i\frac{\erw{c,c'}}{4\pi}} \cdot e^{\frac1{4\pi^2}\big(\wt H(\ell+\ell';c,c')+\wt
  H(\ell+\ell';c',c)-\wt H(\ell-\ell';c,c')-\wt H(\ell-\ell';c',c)\big)}.
  \eea
The second factor is,  in general, 
  not a complex phase. Replacing $c$ by $-c$, if necessary, its
  modulus may be $>1$.
\end{lemma}
 {\em Sketch of proof}: 
By the structure \eref{VVV} of vertex operator correlations,  the four-point function is
a product of six factors. One can convince oneself that in the limit
the smearing functions in $x$ can be neglected, so that two of the six
factors in the numerator cancel against the denominator. This 
essentially leaves the four factors
\bea{four} 
\frac{e^{-\frac{1}{4\pi^2}\wt H(x+t\ell-x'-t\ell';c,c')}}{(-(x+t\ell-x'-t\ell'-i\eps u)^2)^{-\frac{\erw{c,c'}}{8\pi^2}}}\cdot
\frac{e^{-\frac{1}{4\pi^2}\wt H(x'-t\ell'-x+t\ell;c',c)}}{(-(x'-t\ell'-x+t\ell-i\eps
  u)^2)^{-\frac{\erw{c,c'}}{8\pi^2}}}\cdot \notag
\\
\frac{e^{\frac{1}{4\pi^2}\wt H(x+t\ell-x'+t\ell';c,c')}}{(-(x+t\ell-x'+t\ell'-i\eps u)^2)^{+\frac{\erw{c,c'}}{8\pi^2}}} \cdot
\frac{e^{\frac{1}{4\pi^2}\wt H(x'+t\ell'-x+t\ell;c',c)}}{(-(x'+t\ell'-x+t\ell-i\eps u)^2)^{+\frac{\erw{c,c'}}{8\pi^2}}} .
\eea
Because $\ell\pm \ell'$ are timelike resp.\ spacelike, the
denominators in \eref{four} are dominated by powers
of $-t^2((\ell\pm \ell')-i\eps u)^2 = 2t^2[\mp(\ell\ell')
+i\eps(u(\ell\pm\ell')) $. Since $(\ell\ell')>0$, the spacelike cases in the
first line give $[2t^2(\ell\ell')]^{\frac{\erw{c,c'}}{8\pi^2}}$
each. The timelike cases in the second line give
$[2t^2(\ell\ell')(-1+i\eps)]^{-\frac{\erw{c,c'}}{8\pi^2}}$ each. Together,
they yield the first factor in \eref{limit}.
Because $H$ is a homogeneous function, the numerators  in \eref{four}
yield the second factor in \eref{limit}. Now, one may regard
 $L_\pm=\ell\pm\ell'$ as a pair of  orthogonal
 vectors, one timelike and one spacelike, and otherwise
 independent. For $L_-$ purely spacelike, $\wt H(L_-;c,c')+\wt
 H(L_-;c',c)$ is a  real function with a nontrivial dependence on $L_-$. It cannot be cancelled
 by the real part of $\wt H(L_+;c,c')+\wt H(L_+;c',c)$. \qed 

The first factor in \eref{limit} is a phase, as in the two-dimensional
model \cite{DM}.
But because the second factor may have modulus $>1$, \eref{limit}
cannot be interpreted as an S-matrix element.  The challenge remains
to understand
with which modification \eref{tlim} would possibly define an S-matrix,
or whether a variant of the more complicated strategy in \cite{Bu1b},
that was formulated for massless particles in 4D,
should be used. 

 The conclusion is only avoided if all strings are orthogonal to both
$\ell$ and $\ell'$, in which case $H(\ell\pm \ell';e_1,e_2)$ are independent
of $\ell,\ell'$. Namely, in this case $\zeta(\ell\pm\ell';e_1,e_2)$
are independent of $\ell$, $\ell'$.
But by \rref{r:smear}, this would require a
smearing in the intersection of the sphere $S^2$ with a plane (i.e., a
circle $S^1$), which would
in turn render, e.g., the exponents \eref{cc} ill-defined.

\subsection{Derivative formula}
\label{s:deri}

The derivative of \eref{IvW} w.r.t.\ $x$ is surprisingly simple to compute. 
\begin{lemma}\label{l:deri}
  Let
$F(x)=\frac 1{x^2}$. For $x^2$, $e_i^2$, $\det_{e_1,e_2}$,
$\det_{x,e_i}$ and $\det_{x,e_1,e_2}$ all non-zero, it holds
\bea{dIIF}
P:= (I_{e_2}I_{e_1}\pa F)(x) = \frac 12 \big[ f(e_1,e_2)\pa_x +
f(x,e_2)\pa_{e_1} + f(x,e_1)\pa_{e_2}\big] \log
\det{}_{x,e_1,e_2}.
\eea
\end{lemma}
{\em Proof:} 
By definition as a convergent integral over the derivative of $\frac
1{(x+s_1e_1+s_2e_2)^2}$, $P$
is a linear combination $b_1e_1+b_2e_2+b_3x$. From \eref{edI} we know that
$(e_iP)=f(x,e_j)$ for $i,j\in\{1,2\}$ pairwise distinct. From \eref{IvW} we know
that $(xP)= f(e_1,e_2)$. Then
\bea{PG}\bpm(e_1P)\\(e_2P)\\(xP)\epm = \bpm f(x,e_2)\\ f(x,e_1)\\
f(e_1,e_2)\epm = G\bpm b_1\\ b_2\\ b_3\epm,\eea
where $G$ is the Gram matrix,
is solved for the coefficient functions $b_i$ by using \lref{l:Ginv}. \qed

A remarkable feature of \eref{dIIF} is its formal symmetry 
in the three vector variables, despite the unsymmetric definition. The
symmetry can be understood by a change of variables
$(s_1,s_2)\to (\frac1{s_1},\frac{s_2}{s_1})$ in the defining double string-integral.

Given \lref{l:deri}, to specify $(I_{e_2}I_{e_1}\pa W_0)(x)$ for $e_1,e_2\in H_1$
purely spatial, it suffices to specify
the $i\eps$-prescriptions at the possibly singular configurations.
  For $e_1,e_2\in H_1$ purely spatial and $x$ replaced by $x-i\eps u$,
  the distributions $f(x,e_i)$ and $f(e_1,e_2)$ are
  well-defined. It remains to consider the reciprocal of
  $\det_{x,e_1,e_2}$ in \eref{dIIF} when
  $\det_{x,e_1,e_2}=0$. Because $i\eps$ appears only in the term
  $(x-i\eps u)^2\det_{e_1,e_2}$, the distribution is well-defined
  unless $\det_{e_1,e_2}=0$, i.e., when $e_1\neq \pm e_2$. The case
  $e_1=e_2$ is regular by \pref{p:H}. Thus we conclude

\begin{prop}\label{p:deri}
\bea{dIIW}
-(2\pi)^2(I_{e_2}I_{e_1}\pa W_0)(x) = \frac 12 \big[ f(e_1,e_2)\pa_x +
f(x,e_2)\pa_{e_1} + f(x,e_1)\pa_{e_2}\big] \log\det{}_{x,e_1,e_2},
\eea
with the distributions $f(x,e_i)$ as specified in \sref{s:2pt-one} and
$f(e_1,e_2)$ given by \eref{fee}, is well-defined as a distribution on
$\RR^4\times S^2\times S^2\setminus\{(x,e_1,e_2):e_1=-e_2\}$. The
singularity at $e_1=-e_2$ is integrable w.r.t.\ the invariant measure
on $S^2\times S^2$.
\end{prop}

The result can also be obtained with a cumbersome computation of the
derivative of $(I_{e_2}I_{e_1}\pa W_0)(x)$ given by \eref{IvW} and
\eref{H}. This may be taken as a nontrivial check of \eref{H}.

\section{Propagator}
\label{s:propa}
While the IR-regularized {\em two-point function} of the escort field is
needed for the non-perturbative construction of the dressed Dirac field in \cite{MRS3},
{\em propagators} are needed in perturbation theory. Defining QED as a
perturbative expansion with interaction density $q\,A_\mu(e)j^\mu$ in
the Wigner space,
where the two-point function of $A_\mu(e)$ is
\bea{AA}\erw{A_\mu(x,e)A_\nu(x',e')}=-\big[\eta_{\mu\nu} + e_\nu I^x_e\pa^x_\mu + e'_\mu
I^{x'}_{e'}\pa^{x'}_\nu + (ee')I^{x'}_{e'}I^x_e\pa^x_\mu\pa^{x'}_\nu\big] W_0(x-x'),
\eea
one needs only string-integrations over {\em derivatives} of the massless
two-point function and propagator. These are IR-regular. We therefore
focus on the analog of \eref{dIIW} for the propagator.

At first sight, the case of the propagator should be very parallel to
that of the two-point function
except for a different $i\eps$-prescription: $x^2-i\eps$ rather than
$(x-i\eps u)^2$. Instead of an ``analytic continuation through
the forward tube'', one needs an analytic continuation
in the variable $x^2$.  For the single string-integral of the
propagator, it suffices to define the derivative of $f(x,e)$ by simply
substituting $(x-i\eps u)^2$ in \eref{fxe} and \eref{detxe} by
$x^2-i\eps$. 

For the double string-integral over the derivative, however, this
prescription is not sufficient because the reciprocal of $\det_{x,e_1,e_2}$
becomes
\bea{detinv?}\frac1{\det_{x,e_1,e_2} - i\eps\det_{e_1,e_2}}.\eea
In the Wigner space approach to QED, the strings are not restricted to
be purely spatial \cite{MRS3}. But in Lorentzian metric, if $e_i\in H_1$ are not purely spatial, $\det_{e_1,e_2}$ may vanish in a larger submanifold of
configurations, see \lref{l:det20}. For such $e_1,e_2$, \eref{detinv?}
does not define a distribution in $x$ at $\det_{x,e_1,e_2}=0$.
The following \lref{l:signs}, which is a
corollary to \lref{l:MM} allows to reduce the set of
configurations where \eref{detinv?} is not defined.
  \begin{lemma}\label{l:signs} 
    Suppose that $\det_{y_1,y_2,y_3}=0$. Then, for $i,j,k\in\{1,2,3\}$ pairwise
    distinct, hold:\\
    (i) All ``diagonal'' cofactors 
    $\det_{y_j,y_k}$ have the same sign, or vanish.
    \\ (ii) If $\det_{y_i,y_j}=0$, then also the cofactors
    $\Lambda_i$ and $\Lambda_j$ vanish. 
\end{lemma}
{\em Proof:} (i) By \lref{l:MM}, if $\det_{y_1,y_2,y_3}=0$, then
$\det{}_{y_i,y_j}\det{}_{y_i,y_k}=\Lambda_i^2\geq 0$. Thus $\det_{y_i,y_j}$ and $\det_{y_i,y_k}$ cannot be non-zero with
opposite sign.  \\ (ii) is obvious from \lref{l:MM}.
\qed

Because by \lref{l:signs}(i) the three 
two-variable Gram determinants cannot have opposite signs when
$\det_{x,e_1,e_2}=0$, one may define the reciprocal of $\det_{x,e_1,e_2}$ as a pullback of a boundary value of an analytic function
in the variables $x^2-i\eps$, $e_1^2-i\eps$, and $e_2^2-i\eps$, i.e.,
\bea{detinv}\frac1{\det_{x,e_1,e_2} - i\eps(\det_{e_1,e_2}+\det_{x,e_1}+\det_{x,e_2})}.\eea
This prescription, applied to \eref{dIIF},
defines the double string-integral over the derivative of the propagator as a distribution everywhere on
$\RR\times H_1\times H_1$, 
except on configurations where all four Gram determinants vanish
simultaneously. When this happens, then by  \lref{l:MM} (or \lref{l:signs}(ii)), also the
non-diagonal cofactors $\Lambda_i$ of the Gram determinant vanish.

The following Lemmas allow to characterize configurations with vanishing Gram
determinants in Lorentzian metric, and in particular the configurations where \eref{detinv}
is not defined.
\begin{lemma}\label{l:det30}
Suppose that $\det_{y_1,y_2}\neq0$. Then $\det_{y_1,y_2,y_3}=0$ if
and only if there exist $\alpha$, $\beta\in\RR$ and $\ell$ lightlike
($\ell^2=0$) such that 
\bea{Det0}y_3=\alpha y_1+\beta y_2+ \ell \quad\hbox{and}\quad (\ell y_1)=(\ell
y_2)=0.\eea
\end{lemma}
{\em Proof:} Because $\det{}_{y_1,y_2}\neq 0$, one may define $\alpha$, $\beta$ by 
\bea{Det0pf}\bpm y_1^2& (y_1y_2) \\ (y_2y_1) & y_2^2\epm \bpm \alpha\\
\beta\epm = \bpm (y_3y_1) \\ (y_3y_2)\epm.\eea
With $\ell:= y_3-\alpha y_1-\beta y_2$, an elementary computation
gives
\bea{y1y}
\det{}_{y_1,y_2,y_3} = \det{}_{y_1,y_2,\ell}=\ell^2 \det{}_{y_1,y_2}.\eea
Hence $\det{}_{y_1,y_2,y_3}=0$ if and only if $\ell^2=0$.\qed

Analogous statements with analogous proofs hold for $n\times n$ Gram
determinants (which are trivially zero in 4 dimensions for $n>4$).
The version for
$n=2$ is
\begin{lemma}\label{l:det20}
Suppose that $y_1^2\neq0$. Then $\det_{y_1,y_2}=0$ if
and only if there exists $\alpha\in\RR$ and $\ell$ lightlike
such that 
\bea{Det0}y_2=\alpha y_1+ \ell \quad\hbox{and}\quad (\ell y_1)=0.\eea
\end{lemma}

\begin{lemma}\label{l:lindep}
Let $y_1^2\neq0$, $y_2^2\neq0$. If
$\det_{y_1,y_2,y_3}=\det_{y_1,y_2}=0$ and either $\det_{y_1,y_3}=0$ or
$\det_{y_2,y_3}=0$, then the set
$\{y_1,y_2,y_3\}$ is linearly dependent. 
  \end{lemma}
  {\em Proof:} Assume $\det_{y_1,y_3}=0$. Define
$$v:= y_2-\frac{(y_1y_2)}{y_1^2}y_1, \qquad w:=
y_3-\frac{(y_2y_3)}{y_2^2}y_2.$$
One computes that $v^2=w^2=0$ because
$\det_{y_1,y_2}=\det_{y_1,y_3}=0$, and $(vw)=0$ because $\Lambda_2=0$
by \lref{l:signs}(ii). Because two lightlike vectors that are orthogonal must
be linearly dependent, it follows that $v$ and $w$, and consequently
$y_1,y_2,y_3$ are linearly dependent.   The case with $\det_{y_2,y_3}=0$ is similar, replacing
  $y_1\leftrightarrow y_2$. 
  \qed

The conclusion of  \lref{l:lindep} is {\em not} true if
$\det_{y_1,y_2}\neq0$ but the other two $2\times2$ determinants are
zero. It is easy to find counter examples.

We can now classify the configurations where the inverse Gram determinant is not
defined by the prescription \eref{detinv}.
\begin{lemma}\label{l:dets=0}
For 
$(x,e_1,e_2)\in\RR^4\times H_1\times H_1$ it holds 
$\det_{x,e_1,e_2}=\det_{x,e_1}=\det_{x,e_2}=\det_{e_1,e_2}=0$ if and
only if either
\bea{config1}e_1=\pm e_2, \qquad x=\alpha e_1 + \ell\eea
for some $\alpha \in\RR$ and lightlike $\ell$ with $(e_1\ell)=0$, or
\bea{config2} e_2=a'e_1+\ell', \qquad x=\alpha_1e_1+\alpha_2e_2\eea
for some $\alpha',\alpha_1,\alpha_2\in\RR$ and lightlike $\ell'$ with $(e_1\ell')=(e_2\ell')=0$. In
either case, the set $\{x,e_1,e_2\}$ is linearly dependent.
\end{lemma}
{\em Proof:} The ``If'' statements are trivial by linear dependency. Conversely,
$\det_{e_1,e_2}=0$ implies $e_2=ae_1+\ell'$ with $\ell'^2=0$ by
\lref{l:det20}. It then trivially follows that also $(e_2\ell')=0$. If $\ell=0$, then $a=\pm 1$, and \lref{l:det20}
applied to $\det_{x,e_i}=0$ entails \eref{config1}. If $\ell'\neq 0$,
then $e_1$ and $e_2$ are not linearly dependent. Because the three
vectors $x,e_1,e_2$ are inearly dependent by \lref{l:lindep}, $x$ must
be a linear combination of $e_1$ and $e_2$. This is \eref{config2}. \qed

The configurations of type \eref{config1} and \eref{config2} both have codimension 3 in
$\RR^4\times H_1\times H_1$. At this moment, we do not know how to
naturally extend the propagator to these sets. See, however, \cite{G}
where it was proven with microlocal methods that the string-localized
propagator of $A_\mu(x,e)$, when smeared in the strings, can be
defined on all of $\RR^4$. This suggests that one can find coordinates in $$\RR^4\times H_1\times H_1$$ with respect to which the singularities of the inverse determinant at configurations as in \lref{l:dets=0} are integrable, similarly to the integrability of $\frac\alpha{\sin\alpha}$ observed earlier. In \cite{G}, it is also shown that products of smeared string-localized
propagators have natural extensions as distributions in $x$ in $\RR^4\setminus\{0\}$, exactly as in
point-localized QFT.

\bigskip

{\bf Acknowledgements.} We thank J. Mund for valuable discussions. CG thanks
the Studien\-stiftung des deutschen Volkes for financial support.

\edo

\subsection{One string integration: Cuts and branches}
\label{s:onedist}
To deal with the distribution $(I_eW_0)(x)$ on the entire
configuration space, it suffices to recall \eref{W0} that represents
$W_0(x)$ as the boundary value of an analytic function in the
complex tube
$\IM(x^0)<0$. In other words, one may just replace $x$ by $x-i\eps n_0$
where $n_0$ is a future timelike vector. $n_0$ may be chosen orthogonal
to $e$ so that $(xe)$ is unchanged, while $\det_{x,e}$ is replaced by
$\det_{x,e}+i\eps x^0$, where $x^0=(xn_0)$. Then the distribution
$$(I_eW_0)(x) 
= \frac{1}{(2\pi)^2}
\lim_{\eps\downarrow0} \int_{-(xe)}^\infty
\frac{du}{u^2+\det_{x,e}+i\eps x^0}  =
\frac{1}{(2\pi)^2}
\lim_{\eps\downarrow0}\frac{\frac12\log\frac{-(xe)+\sqrt{-\det_{x,e}-i\eps
    x^0}}{-(xe)-\sqrt{-\det_{x,e}-i\eps x^0}}}{\sqrt{-\det_{x,e}-i\eps
x^0}}
$$
is given in the cases (D--E) by analytic continuation of \eref{I1reg}, 
which is defined in the forward tube of the regular points, to the forward tube
of the singular points.
In the same way, by \eref{G0F},
$$(I_eG_0^F)(x) 
= \frac{i}{(2\pi)^2}
\lim_{\eps\downarrow0} \int_{-(xe)}^\infty
\frac{dt}{t^2+\det_{x,e}+i\eps} =
\frac{i}{(2\pi)^2}
\lim_{\eps\downarrow0}\frac{\frac12\log\frac{-(xe)+\sqrt{-\det_{x,e}-i\eps}}{-(xe)-\sqrt{-\det_{x,e}-i\eps}}}{\sqrt{-\det_{x,e}-i\eps}}.
$$
\blue{Man könnte sie
  noch ausrechnen in den Gebieten D und E: dort ist $\det_{x,e}$
  negativ, also 
  $\sqrt{-\det_{x,e}-i\eps} = \sqrt{-\det_{x,e}}-i\eps$. In D ist der
  Zähler im Logarithmus positiv und der Nenner negativ; in E sind
  Zähler und Nenner negativ. Benutze jeweils $\log(-z\pm i\eps)= \pm
  i\pi+\log z$ für $z>0$. Also
$$(I_eG_0^F)(x) =
\frac{i}{(2\pi)^2}
\lim_{\eps\downarrow0}\frac{\frac12\log\Big\vert\frac{-(xe)+\sqrt{-\det_{x,e}}}{-(xe)-\sqrt{-\det_{x,e}}}\Big\vert
  -\frac n2i\pi}{\sqrt{-\det_{x,e}}}$$
mit $n=1$ in D und $n=2$ in E; ebenso aber mit $\frac
n2i\pi\cdot\sign(x^0)$ für $(I_eW_0)(x)$.}

Because $(e\pa)\det_{x,e}=0$, the application of $-(e\pa)$ to these
distributions hits only the lower integration limit $-(xe)$, and thus
reproduces the distributions $W_0(x)$ and $G_0^F(x)$.

\subsection{Two string integrations: a priori considerations}
\label{apriori}

Because the two-point function $W_0(x)$ is homogeneous of degree $-2$,
$(I_eW_0)(x)$ is homogeneous of degree $-1$, and
\bea{IaW}(I^a_{-e'}I_eW_0)(x):= \int_0^a ds' (I_eW_0)(x-s'e') \equiv
\int_0^a ds' \ioi ds\, W_0(x+se-s'e')\eea
diverges logarithmically as $a\to\infty$. In momentum space,
$$(I_{-e'}I_e W_0)(x):=\lim_{\eps\downarrow0}\lim_{\eps'\downarrow0}\int d\mu_0(k)
\frac{e^{-ikx}}{((ke)-i\eps)((ke')+i\eps')}$$
diverges at $k=0$. We therefore have to regularize it in the
infrared. This is done by replacing $e^{-ikx}$ by $e^{-ikx}-v(k)$ where $v(k)$
is any smooth test function with $v(0)=1$. Thus, we define
\bea{IvWdef}(I_{-e'}I_e W_0)_v(x):=\lim_{\eps\downarrow0}\lim_{\eps'\downarrow0}\int d\mu_0(k)
\frac{e^{-ikx}-v(k)}{((ke)-i\eps)((ke')+i\eps')}.\eea
For the Feynman propagator, one only has to
replace in the Fourier representation the measure
where $d\mu_0(k)=\frac1{(2\pi)^3}\,
d^4k\,\delta(k^2)\theta(k^0)=\frac1{(2\pi)^3}\frac{d^3k}{2\vert\vec k\vert}$.
 by $\frac
 {-1}{(2\pi)^4}\frac{d^4k}{k^2+i\eps}$.
 We want to gain insight into the distribution $(I_{-e'}I_e W_0)_v(x)$
 in position space, leaving the regulator function $v$ unspecified.
 
 By \eref{IvWdef}, 
 $$\pa_\mu (I_{-e'}I_e W_0)_v(x) = (I_{-e'}I_e \pa_\mu W_0)(x)$$
 is independent of $v$, and can be computed from the cut-off integral \eref{IaW}: 
\bea{dIvdIa}\pa_\mu (I_{-e'}I_e W_0)_v(x) = \lim_{a\to\infty}(I^a_{-e'}I_e
\pa_\mu W_0)(x)= \lim_{a\to\infty}\pa_\mu(I^a_{-e'}I_e W_0)(x).\eea
We shall compute below
$$(I^a_{-e'}I_e W_0)(x) =\frac{-1}{(2\pi)^2} \Big[-\frac12 f(e,-e')\cdot
\log\big(\frac{4(ae')^2}{(x^2)_-}\big) + \frac{H(x;e,-e')}{(ee')}\Big]$$
where $f(e,-e')=f(-e',e)$ is the same distribution as in \eref{I1s}, and
the distribution $H$ is symmetric in $e\lra -e'$ and homogeneous of
degree $0$ in $x$ and in $e$ and in $e'$. Then, by
\eref{dIvdIa},
$$(x\pa)(I_{-e'}I_e W_0)_v(x) = \frac{-1}{(2\pi)^2} \cdot \frac12 f(e,-e')$$
is independent of $x$, and
\bea{IvW}(I_{-e'}I_e W_0)_v(x) = \frac{-1}{(2\pi)^2} \Big[\frac12f(e,-e')\cdot \log(-\mu^2_v(e,-e')^2(x^2)_-) +
\frac{H(x;e,-e')}{(ee')}\Big]\eea
where $\mu_v$ is independent of $x$ and carries the dependence on $v$.
$\mu_v$ may depend on $e,e'$. If it does, it is symmetric in $e\lra -e'$ and homogeneous of
degree $0$ in $e$ and in $e'$.

The same holds with the obvious modifications for $(I_{-e'}I_eG^F)_v(x)$. 

The logarithmic behaviour of regularized string-localized correlation functions and
propagators is controlled by the function $f(e,-e')$, e.g., for the
escort field
$$\omega_v(\varphi(x_1,c_1)\varphi(x_2,c_2)) = -\frac{\erw{c_1,c_2}}{8\pi^2}\cdot \log(-\mu_v^2x^2)
+ \hbox{homogeneous}.$$
where $\erw{\cdot,\cdot}$ is an inner product in $C^\infty(S^2)$ whose
kernel is given by the function $f$, see \sref{s:vertex}.
For the exponentiated escort fields as outlined in the introduction,
the $\log(x^2)$-term becomes a power law in $x^2$, while the
homogeneous term provides directional ``modulations''.

\subsection{Two string integrations: Regular configurations}
\label{s:tworeg}

In the sequel, in order to better exhibit the symmetry $e\lra -e'$
(apart from the asymmetric integration $I_{-e'}^aI_e$), we relabel
$e=e_1$ and $-e'=e_2$.

We start again with the regular configurations where the ``quarter
plane'' $x+\RR_+e_1+\RR_+e_2$ does not intersect the null-cone. These
include the ``purely spatial configurations'' with $x^0=e_1^0=e_2^0=0$
such that $0\notin x+\RR_+e_1+\RR_+e_2$.

We compute for $F(x)=\frac1{x^2}$
\bea{IaI}(I^a_{e_2}I_{e_1}F)(x) := \int_0^a ds_2 (I_{e_1}F)(x+s_2e_2)= -\int_0^a ds_2
\,f(x+s_2e_2,e_1).\eea
This integral is finite for finite $a$ and diverges logarithmically
when $a\to\infty$. 

{\bf The case of general regular configurations}

The second string integral is computed in the appendix \sref{a:splike}. A crucial
quantity is the determinant
$$\textstyle{\det_{x,e_1,e_2}}:=x^2e_1^2e_2^2-x^2(e_1e_2)^2-e_1^2(xe_2)^2-e_2^2(xe_1)^2+2(xe_1)(xe_2)(e_1e_2),$$
whose vanishing in the purely spatial case signals that $x$, $e_1$ and
$e_2$ are linearly dependent, ie, the plane spanned by $e_1$ and $e_2$
emanating from $x$ contains the origin.

It is convenient to introduce the
homogeneous and Lorentz invariant coordinates
$$\framebox{$\displaystyle\quad u_k=\frac{(xe_k)}{\sqrt{x^2e_k^2}}
  \qquad (k=1,2),\qquad
\qquad v=\frac{(e_1e_2)}{\sqrt{e_1^2e_2^2}}.\quad$}$$
For purely spatial configurations $x=(0,\vec x)\neq 0$, $e_k=(0,\vec e_k)$,
and $$\alpha_k=\angle(\vec x,\vec e_k)=\pi-\beta_k\qquad (k=1,2), \qquad
\qquad \gamma =\angle(\vec e_1,\vec
e_2)=\pi-\theta,$$
these are
$$u_k=-\cos\alpha_k=\cos\beta_k\qquad (k=1,2),\qquad\qquad v=-\cos\gamma=\cos\theta.$$
The configurations $\gamma=\pi$ ($e_1=-e_2$, where $I_{e_2}I_{e_1}$ is
ill-defined even without the logarithmic divergence) and $\alpha_1+\alpha_2+\gamma=2\pi$ (there exist $s_1\geq0$ and
$s_2\geq0$ such that $x+s_1e_1+s_2e_2=0$; including $\alpha_1=\pi$ or
$\alpha_2=\pi$) are possibly singular; all other purely
spatial configurations are regular. In these configurations,
$$\sqrt{e_1^2e_2^2}\cdot f(e_1,e_2) = \frac{\frac1{2i}\log\frac{-v+ i\sqrt{1-v^2}}{-v-
    i\sqrt{1-v^2}}}{\sqrt{1-v^2}} =
\frac{\frac12\log\frac{e^{i\gamma}}{e^{-i\gamma}}}{
  i\sin\gamma} = \frac{\gamma}{\sin\gamma}$$
(which is regular at $\gamma=0$, but singular at $\gamma=\pi$, as
expected). 

Similarly, 
$$f(u_k):=\sqrt{x^2e_k^2}\cdot f(x,e_k) = \frac{\frac1{2i}\log\frac{-u_k+ i\sqrt{1-u_k^2}}{-u_k-
    i\sqrt{1-u_k^2}}}{\sqrt{1-u_k^2}} \qquad (k=1,2),$$
such that in accord with \eref{fxe-spl}
$$ f(u_k) = \frac{\alpha_k}{\sin(\alpha_k)}.$$
Let further\footnote{in $D$ hat christian gegenüber den hier verwendeten konventionen das umgekehrte vorzeichen im letzten
  term $2u_1u_2v$, und in den $A$'s das umgekehrte vorzeichen im linearen
  term $u_2$ bzw $u_1$. }

\begin{minipage}{97mm}$$D:=\frac{\det_{xe_1e_2}}{x^2e_1^2e_2^2} =
  1-u_1^2-u_2^2-v^2-2u_1u_2v \geq0$$
  which vanishes whenever $x+s_1e_1+s_2e_2=0$ for some $s_1,s_2>0$,
  i.e., when $\alpha_1+\alpha_2+\gamma=2\pi$ ($\Leftrightarrow$
  $\beta_1+\beta_2+\theta=\pi$). 
  \end{minipage} \begin{minipage}{60mm}
$$\includegraphics[width=40mm]{triangle.pdf}$$
\end{minipage}
In the sequel, let always $j=2$ if $k=1$ and vice versa.
Define
$$A_k^\pm :=
u_kv+u_j\pm\sqrt{(1-v^2)(1-u_k^2)}. $$
Then $A_k^+>0$ and $A_k^-<0$, and
\bea{ApAm}A_k^+A_k^-  = -D.\eea
It is therefore also convenient to write
\bea{zeta}\frac{A_k^\pm}{\sqrt D}= \pm e^{\pm \zeta_k}.
\eea
The actual computation of $(I^a_{e_2}I_{e_1}F)(x)$
is presented in \aref{a:twostrings}.\footnote{die rohformel (1.1) in hom-coord
  hat in der ersten zeile den term $f\log(a\Gamma_1) =f\log\wt a
  =-\frac12 f\log(x^2)$. das scheint das falsche vorzeichen zu
  sein. das ist aber
  nicht so, weil ein divergenter beitrag von der oberen grenze genau
  das $-2$-fache produziert, also das vorzeichen umkehrt.} It is of the form announced in \sref{apriori}. In terms of the homogeneous
coordinates, it is
\begin{framed}
  $$
  \sqrt{e_1^2e_2^2}(I^a_{e_2}I_{e_1}F)(x) = -\frac12\frac\gamma{\sin(\gamma)}
  \log\big(\frac{4(ae_2)^2}{x^2}\big)-
  \frac\gamma{\sin(\gamma)}\log\big(\frac{1-v^2}{\sqrt D}\big)-\hspace{40mm}$$
\vskip-6mm
\bea{IaFtotal}\hspace{10mm}-
\frac{\pi}{2\sin(\gamma)}\log\big(\frac{A_1^+A_2^+}D\big)+\frac i{4\sin(\gamma)}\left\{\Li_2\Big(e^{i\gamma}\frac{A_1^+A_2^+}{D}\Big)\ba{c}+(A_1^+\lra
A_1^-)\\ +(A_2^+\lra
A_2^-)\\-(e^{i\gamma}\lra e^{-i\gamma})\ea\right\} + O(\frac1a).\quad
\eea
\end{framed}
By its manifest symmetries, this formula is independent of all choices of
square roots $\sqrt{-x^2}$ (that would simultaneously revert both $u_k$),
$\sqrt{1-v^2}$ (that would revert $\gamma$),
$\sqrt{1-u_k^2}$ (that would swap $A_k^+\lra A_k^-$). 
The logarithmic dependence on $-x^2$ is consistent with
\eref{e-e'} and \eref{e-e's} at $e_1=e_2$ where
$\frac\gamma{\sin(\gamma)}=f(e_1,e_1)=1$. 

In terms of the $\zeta$-variables \eref{zeta},
$\log\big(\frac{A_1^+A_2^+}{D}\big) = \zeta_1+\zeta_2$, and the dilog
sum in \eref{IaFtotal} becomes
\bea{lizeta}\Big\{\Li_2\big(e^{i\gamma}e^{\zeta_1+\zeta_2}\big)+\Li_2\big(e^{i\gamma}e^{-\zeta_1-\zeta_2}\big)+\Li_2\big(-e^{i\gamma}e^{-\zeta_1+\zeta_2}\big)+\Li_2\big(-e^{i\gamma}e^{\zeta_1-\zeta_2}\big)\Big\}-(e^{i\gamma}\lra
e^{-i\gamma}).\quad \eea

In the limit $\gamma\to 0$ ($
e_1\to e_2$), the $x$-independent term $\log(1-v^2)\sim2\log(\gamma)$ secures the absence of a singularity
$\sim\log(\gamma)$ that is present both in $\log(\sqrt D)\sim\log(\gamma)$ and
in the dilog sum $\sim-4i\gamma\log(\gamma)$ (see \sref{a:splike}).
In the appendix, we compute the regular limit 
at $e_1=e_2$, and check that it coincides with the known result \eref{e-e'}. In
contrast, at $\gamma\to\pi$ ($\theta=\pi-\gamma\to0$, $e_1\to -e_2$) one
finds on top of the integrable singularity
$\frac\gamma{\tan(\gamma)}\sim -\frac\pi{\tan(\theta)}$, as in
\sref{s:onereg}, logarithmic
singularities like $\log(\theta)/\theta$ that are integrable
as well.

\subsection{Two string integrations: Cuts and branches}
\label{s:twocuts}

\blue{Ähnlich wie \sref{s:onedist}}

In the purely spatial situations, all square roots are real, and are 
understood as the positive ones. The $i\eps$ prescriptions will then
control the extension to singular configurations.

TODO

For spacelike configurations $x=(t,r\vec n)$ where $r>\vert t\vert$, $e_k=(0,\vec e_k)$,
one finds that the strings emanating from $x$
touch the lightcone iff one of the $u_k=1$ such that $A_j^+=A_j^-=u_k+v$.
When they cut through the
lightcone, then $u_k>1$, and $A_k^\pm$ become complex.

\subsection{Vertex operators}
\label{s:vertex}
The 2-pt fn of the escort field is the IR-regulated
version of \eref{IaFtotal} with $e_1=e=(0,\vec e)$ and
$e_2=-e'=-(0,\vec e\,')$, multiplied by
$\frac{(ee')}{(2\pi)^2}=-\frac{\cos\theta}{(2\pi)^2}$, where $\theta = \theta_{\vec e,\vec e\,'}=\pi-\gamma$: 
$$\erw{\phi(x,e)\phi(x',e')}_v =
-\frac1{8\pi^2}\frac{\pi-\theta}{\tan\theta}
 \Big[ \log\big(-\mu_v^2x^2\big)-2\log\big(\frac{1-v^2}{\sqrt D}\big)\Big]+
\hspace{40mm}$$
\vskip-6mm
$$+\frac{\pi}{8\pi^2\tan\theta}\log\big(\frac{A_1^+A_2^+}D\big)-\frac i{16\pi^2\tan\theta}\left\{\Li_2\Big(e^{i(\pi-\theta)}\frac{A_1^+A_2^+}{D}\Big)\ba{c}+(A_1^+\lra
A_1^-)\\ +(A_2^+\lra
A_2^-)\\-(e^{i(\pi-\theta)}\lra e^{-i(\pi-\theta)})\ea\right\}
$$
Thus, 
$$\Erw{\wick{e^{iq\phi(x,c)}}\wick{e^{-iq\phi(x',c')}}}_v =\delta_{c,c'}\cdot
e^{q^2\erw{\phi(x,c)\phi(x',c')}_v} \sim \delta_{c,c'}\cdot
W_0(x-x')^{\frac{q^2}{8\pi^2}\cdot \erw{c,c}},$$
where $\sim$ stands for a homogeneous factor, and 
$$\erw{c,c'}:= \int d\sigma(\vec e)c(\vec e)\int d\sigma(\vec
e\,')c(\vec e\,')\,\frac{\pi-\theta_{\vec e,\vec
    e\,'}}{\tan(\theta_{\vec e,\vec e\,'})}.$$
In particular for the constant smearing $c_0(e)=\frac 1{4\pi}$,
$$\erw{c_0,c_0} = \frac{1}{(4\pi)^2}\int d\sigma(\vec e)\int d\sigma(\vec
e\,')\,\frac{\pi-\theta_{\vec e,\vec
    e\,'}}{\tan(\theta_{\vec e,\vec e\,'})} = \frac1{4\pi}
\int d\sigma(\vec
e)\,\frac{\pi-\theta}{\tan\theta} =\frac 12\int_0^\pi \sin\theta\,
d\theta \,\frac{\pi-\theta}{\tan\theta} =$$$$=\frac 12\int_0^\pi 
d\theta \, (\pi-\theta)\cos\theta
=\frac12\Big[(\pi-\theta)\sin\theta\big\vert_{0}^\pi + \int_0^\pi
d\theta \sin\theta \Big]=1. $$
Among all real smearing functions of total weight 1, $c_0$ is a local
minimum of $\erw{c,c}$, and we conjecture that it is also a global
minimum, hence $\erw{c,c}\geq 1$.

For the constant smearing $c_0$, also the homogeneous factor can be computed. In this case, the two-pt fn in
momentum space is
$$\erw{\phi(x,c_0)\phi(x',c_0)}_v=\frac1{(4\pi)^2}\oint d\sigma_0(\vec e)\oint d\sigma_0(\vec e\,')\int
d\mu_0(k) \big[e^{-ik(x-x')}-v(k)\big]\frac{\vec e}{(\vec k\cdot \vec
  e)_+}\cdot\frac{\vec e\,'}{(\vec k\cdot\vec
  e\,')_-}.$$
With $\frac1{4\pi}\oint d\sigma_0(\vec e) \frac{\vec e}{(\vec k\cdot \vec
  e)_\pm} = \frac {\vec k}{\vert \vec k\vert^2}$, this becomes
$$\int d\mu_0(k) \frac{e^{-ik(x-x')}-v(k)}{(k^0)^2}=
-\erw{\phi(x,\pm u_0)\phi(x',\mp u_0)}_v,$$
where $u_0=(1,\vec 0)$ is the time unit vector. This two-pt fn can be
computed as in \sref{a:ee'}, replacing $e=-e'$ by $\pm u_0$. The
result is (SIGN under first $\log$ CORRECTED!
$$(2\pi)^2\erw{\phi(x_1,c_0)\phi(x_2,c_0)}_v = -\frac12\log\big(-\mu_v^2 (x^2-i\eps
x^0)\big) -\frac {x^0}{2r} \log\frac{(x^0-i\eps)+r}{(x^0-i\eps)-r}$$
where $x=x_1-x_2$, hence 
$$\Erw{\wick{e^{iq\phi(x_1,c_0)}}\wick{e^{-iq\phi(x_2,c_0)}}}_v =
\Bigg[\frac
{\Big(\frac{x^0-r-i\eps}{x^0+r-i\eps}\Big)^{\frac{x^0}{
      r}}}{-\mu_v^2(x^2-i\eps x^0)}\Bigg]^\alpha , \qquad
\alpha = \frac{q^2}{8\pi^2}.$$

\newpage

The exponent $\alpha\equiv\frac{q^2}{8\pi^2}\erw{c_0,c_0}$ of the power law will modify the decay of correlations of the
Dirac field in the
LSZ limit. 
The numerator describes homogeneous, but only rotational invariant
modifications of the Lorentz
invariant power law.

\appendix

\section{Two string integrals: detailed calculations}
\label{a:twostrings}

\subsection{The case $e_1=e_2$}
\label{a:ee'}

The case $e_1=e_2$ is particularly simple and can be used to check the
subsequent results for the general case. Namely, because
$\det_{x+se,e}=\det_{x,e}$ in \eref{fxe} does not depend on $s$,
$$(I^a_{e}I_eF)(x) = -\int_0^a ds
\,f(x+se,e) =
\frac{i}{2\sqrt{\det_{x,e}}}\int_0^ads\, \log\frac{-(xe)-se^2+i\sqrt{\det_{x,e}}}{-(xe)-se^2-i\sqrt{\det_{x,e}}}
$$
becomes an elementary integral. The result for large $a$ is
\bea{e-e'}\framebox{$\displaystyle\quad (I^a_{e}I_eF)(x) =
\frac1{2e^2}\log\frac{(ae)^2}{x^2}+\frac{1}{e^2}+\frac{(xe)}{e^2} f(x,e) +O\big(\frac1a\big).\quad$}\eea
We retain this for later reference, eg, checking the general case in
the limit $e_2\to e_1$. Notice that the last term secures that
$-(e\pa)(I^a_{e}I_eF)(x) = (I^a_{e}F)(x) \to (I_{e}F)(x)= -f(x,e)$.

For the sake of comparison, one may also proceed more symmetrically
and regularize the divergent integral
$$(I_e^2F)(x) = \ioi ds_1 \, ds_2 \,F(x+s_1e+s_2e) = \ioi s\,ds
\, F(x+se)$$
with an upper integration limit ${a^*}$:
$$(I_e^2 F)^{{a^*}}(x) := \int_0^{{a^*}} s\, ds \, F(x+se) = \frac 1{e^2}\int_{-(xe)}^{-(xe)-e^2{{a^*}}}
\frac {(u+(xe))\,du}{u^2+\det_{x,e}}.$$
The contribution from the constant term in the numerator is
$$-\frac{(xe)}{e^2} f(x,e) + O\big(\frac1{a^*}\big),$$
and the contribution from the linear term is the elementary
integral
$$\frac 1{2e^2}\log(u^2+\det{}_{x,e})\Big\vert_{-(xe)}^{-(xe)-e^2{a^*}} =
\frac1{2e^2} \log \frac{((xe)+{a^*}e^2)^2+\det_{x,e})}{x^2e^2}\sim \frac
1{2e^2}\log \frac{e^2{a^*}^2}{x^2},$$
that exhibits the logarithmic divergence as $a\to\infty$. Thus,
in accord with \eref{e-e'}
\bea{e-e's}\framebox{$\displaystyle\quad (I_e^2 F)^{{a^*}}(x) =
\frac 1{2e^2}\log\frac{e^2{a^*}^2}{x^2}+\frac{(xe)}{e^2} f(x,e) + O\big(\frac1{a^*}\big).\quad$}\eea

\subsection{The case of purely spatial configurations}
\label{a:splike}

For purely spatial regular configurations, the determinants $\det_{xe_k}$
and $\det_{e_1e_2}$ are positive, while $\det_{xe_1e_2}$
is negative. As a consequence, $u_k$, and $v$ as well as $A_k^\pm$,
$D$ and $\zeta_k$ in \sref{s:tworeg} are real, $D>0$, and $0<\gamma<\pi$. 

Starting from \eref{fxe}, the second string integral can be done (CG,
Feb 25) with the substitution of
variables
$$C(s) = e^{\arcsinh(a\Gamma_1+\Gamma_2)}$$
where
$$\Gamma_1=\frac{\det_{e_1e_2}}{\sqrt{e_1^2\det_{xe_1e_2}}}=\sqrt{\frac{e_2^2}{x^2}}\frac{1-v^2}{\sqrt
  D},\qquad \Gamma_2=\frac{e_1^2(xe_2)-(xe_1)(e_1e_2)}{\sqrt{e_1^2\det_{xe_1e_2}}}=-\frac{u_1v+u_2}{\sqrt
  D}.$$
Then (CG, Feb 25)
\begin{framed}
  $$(I^a_{e_2}I_{e_1}F)(x) = \Big[-\frac
1{\sqrt{\det_{e_1e_2}}}(\frac\pi2-\arctan B) \log(C(s))-\hspace{50mm}
$$$$\hspace{30mm}-\frac
i{2\sqrt{\det_{e_1e_2}}}\Big\{\Li_2\Big(\frac{-B+i}{A-\sqrt{1+A^2+B^2}}C(s)\Big)\ba{c}+(\sqrt{\dots}\lra-\sqrt{\dots})\\-(i\lra
-i)\ea\Big\}\Big]_{s=0}^a,$$
\end{framed}
where $$A=\frac{e_1^2(e_2^2(xe_1)-(e_1e_2)(xe_2))}{\sqrt{\det_{e_1e_2}}\sqrt{e^2\det_{xe_1e_2}}}
= \frac{u_1+u_2v}{\sqrt{1-v^2}\sqrt D},$$
$$B=\frac{(e_1e_2)}{\sqrt{\det_{e_1e_2}}} = \frac v{\sqrt{1-v^2}} =
  \frac{-\cos\gamma}{\sin\gamma} = -\frac{\sin(\frac\pi2-\gamma)}{\cos(\frac\pi2-\gamma)},
  $$
  hence
  $$\frac\pi2-\arctan B=\pi-\gamma,$$
  $$\sqrt{1+A^2+B^2}=\frac{1-u_2^2}D,$$
$$\frac{-B+i}{A-\sqrt{1+A^2+B^2}} = e^{i\gamma}\frac{\sqrt D}{A_2^-} =
-e^{i\gamma}\frac{A_2^+}{\sqrt D}.$$
Thus, with $\sqrt{1-v^2}=\sin(\gamma)$
$$
\framebox{$\displaystyle\quad 
\sqrt{e_1^2e_2^2}(I^a_{e_2}I_{e_1}F)(x) = \Big[\frac
{\gamma-\pi}{\sin(\gamma)} \log(C(s))
-\frac
i{2\sin(\gamma)}\Big\{\Li_2\Big(-e^{i\gamma}\frac{A_2^+}{\sqrt D}C(s)\Big)\ba{c}+(A_2^+\lra
A_2^-)\\-(e^{i\gamma}\lra e^{-i\gamma})\ea\Big\}\Big]_{s=0}^a.\quad$}$$
With $\arcsinh z=\log(z+\sqrt{z^2+1})$,
$$C(0)=e^{\arcsinh \Gamma_2}=\frac {-A_1^-}{\sqrt D},$$
$$C(a)=e^{\arcsinh (a\Gamma_1+\Gamma_2)} =
2\sqrt{\frac{(ae_2)^2}{x^2}}\frac{1-v^2}{\sqrt D} - 2\frac{u_1v+u_2}{\sqrt
  D} + O(\frac 1a),$$
$$
\log(C(a)) =\log\big(2\sqrt{\frac{(ae_2)^2}{x^2}}\frac{1-v^2}{\sqrt D}\big) +
O(\frac 1a).$$

{\bf Upper integration limit}

The identity
\bea{Li1z}
\Li_2(z)=-\Li_2(\frac1z)-\frac{\pi^2}6-\frac12(\log(-z))^2\eea
allows to express $\Li_2(C(a))$ by $\Li_2(\frac1{C(a)})=O(\frac1a)$. Thus, the contribution from the upper integration limit is
$$\frac{\gamma-\pi}{\sin(\gamma)}
\log(C(a))+\frac
1{\sin(\gamma)}
\IM \Big\{-\frac12\big(\log\big(e^{i\gamma}\frac{A_2^+}{\sqrt D}C(a)\big)\big)^2-\frac12\big(\log\big((-e^{i\gamma})\frac{-A_2^-}{\sqrt D}C(a)\big)\big)^2\Big\}+O(\frac1a)=$$
$$=\frac{\gamma-\pi}{\sin(\gamma)}
\log(C(a))
-\frac1{2\sin(\gamma)}\IM
\Big\{\big(i\gamma +\log\big(\frac{A_2^+}{\sqrt D}C(a)\big)\big)^2
+\big(i(\gamma-\pi) +\log\big(\frac{-A_2^-}{\sqrt D}C(a)\big)\big)^2\Big\}+O(\frac1a)=$$
$$=\frac{\gamma-\pi}{\sin(\gamma)}
\log(C(a))
-\frac1{\sin(\gamma)}
\Big\{\gamma\log\big(\frac{A_2^+}{\sqrt D}C(a)\big)
+(\gamma-\pi)\log\big(\frac{-A_2^-}{\sqrt D}C(a)\big)\Big\}+O(\frac1a)=$$
\bea{upper}=-\frac\gamma{\sin(\gamma)}
\log(C(a))
-\frac\gamma{\sin(\gamma)}\log(\frac{-A_2^-A_2^+}D)+
\frac\pi{\sin(\gamma)}\log(\frac{-A_2^-}{\sqrt D})+O(\frac1a),
\eea
where we have used that $A_2^+>0$ and $A_2^-<0$, to
avoid writing logarithms of negative numbers. 
The first term is the expected divergence
$$\sim
-\frac{\gamma}{2\sin(\gamma)}\log\big(\frac{(ae_2)^2}{x^2}\big).$$
The
second term $=0$ by \eref{ApAm}, and the last terms will be dealt with
later.

{\bf Lower integration limit}

The contribution from the lower integration limit is
\bea{lower}
-\frac{\gamma-\pi}{\sin(\gamma)}\log\big(\frac{-A_1^-}{\sqrt D}\big)
-\frac 1{\sin(\gamma)}\IM\Big\{\Li_2\Big(e^{i\gamma}\frac{A_1^-A_2^+}{D}\Big)+(A_2^+\lra
A_2^-)\Big\}.
\eea
We split the dilog sum $\IM\{\dots\}$ into a symmetric and an antisymmetric part wrt
$A_1^-\lra A_1^+$. The symmetric part
$$\frac12\IM\Big\{\Li_2\Big(e^{i\gamma}\frac{A_1^+A_2^+}{D}\Big)\ba{c}+(A_1^+\lra
A_1^-)\\ +(A_2^+\lra
A_2^-)\ea\Big\}$$
is also symmetric under $A_1^\pm\lra A_2^\pm$. The antisymmetric part is
$$\frac12\IM\Big\{\Li_2\Big(e^{i\gamma}\frac{A_1^-A_2^+}{D}\Big)-\Li_2\Big(e^{i\gamma}\frac{A_1^+A_2^+}{D}\Big)+(A_2^+\lra
A_2^-)\Big\}=$$
$$=\frac12\IM\Big\{\Li_2\Big(e^{i\gamma}\frac{A_1^-A_2^+}{D}\Big)+\Li_2\Big(e^{-i\gamma}\frac{A_1^+A_2^-}{D}\Big)+(A_2^+\lra
A_2^-)\Big\}.$$
Because the arguments of the two displayed dilogs are each other's
inverse, we can again use $\Li_2(z)+\Li_2(\frac1z) =
-\frac{\pi^2}6-\frac12(\log(-z))^2$, and get
$$=\frac12\IM\Big\{-\frac12\big(i\gamma+\log\big(\frac{-A_1^-A_2^+}{D}\big)\big)^2
-\frac12\big(i(\gamma-\pi)+\log\big(\frac{A_1^-A_2^-}{D}\big)\big)^2\Big\}
=$$
$$=\frac12\Big\{-\gamma\log\big(\frac{-A_1^-A_2^+}{D}\big)
-(\gamma-\pi)\log\big(\frac{A_1^-A_2^-}{D}\big)\Big\}
=$$
$$=-\frac\gamma2\log\big(\frac{-(A_1^-)^2A_2^+A_2^-}{D^2}\big)
+\frac\pi2 \log\big(\frac{A_1^-A_2^-}{D}\big)=-\frac\gamma2\log\big(\frac{(A_1^-)^2}{D}\big)
+\frac\pi2 \log\big(\frac{A_1^-A_2^-}{D}\big).$$
Multiplied with
$-\frac 1{\sin(\gamma)}$ as in \eref{lower}, this term combines with
the first term in \eref{lower} and the last term \eref{upper}:
$$
-\frac{\gamma-\pi}{\sin(\gamma)}\log\big(\frac{-A_1^-}{\sqrt D}\big)
-\frac 1{\sin(\gamma)}\Big(-\frac\gamma2\log\big(\frac{(A_1^-)^2}{D}\big)
+\frac\pi2
\log\big(\frac{A_1^-A_2^-}{D}\big)\Big)+\frac\pi{\sin(\gamma)}\log(\frac{-A_1^-}{\sqrt
  D})=
$$$$= \frac{\pi}{2\sin(\gamma)}\log\big(\frac{A_1^-A_2^-}{D}\big)= -\frac{\pi}{2\sin(\gamma)}\log\big(\frac{A_1^+A_2^+}{D}\big).
$$
{\bf All terms collected}

Collecting all terms, and representing the result again as the
analytic function that it is, rather
than as the imaginary part of an analytic function (which is correct only when
$u$, $u'$, $v$, and all square roots are real, as is the case in spacelike configurations)
we arrive at \eref{IaFtotal}.

\subsection{Cross checks}
{\bf Derivatives I}

Because the computations leading to \eref{IaFtotal} are quite subtle,
we have subjected the result to various checks. The first check
verifies the conditions
$$\lim_{a\to\infty}(e_k\pa)(I^a_{e_2}I_{e_1}F)(x)=-(I_{e_j}F)(x)=f(x,e_j)
$$
(where $j=2$ if $k=1$ and vice versa) that reflect the property that $I_e$ are inverted by
$-(e\pa)$. In terms of the homogeneous
coordinates, these equations are equivalent
to
\bea{paH}\pa_{u_k} H(u_1,u_2,v)= - \frac\gamma{\tan(\gamma)}\frac{u_jv+u_k}D +
\frac vD\Big[(u_1u_2+v)f(u_k)+(1-u_j^2)f(u_j)\Big],\eea
where
\bea{H}H(x,e_1,e_2)=\hspace{130mm}\eea
\vskip-10mm
$$=-\frac\gamma{\tan(\gamma)}\log\big(\frac{1-v^2}{\sqrt D}\big)-
\frac{\pi}{2\tan(\gamma)}\log\big(\frac{A_1^+A_2^+}{D}\big)
+\frac i{4\tan(\gamma)}\left\{\Li_2\Big(e^{i\gamma}\frac{A_1^+A_2^+}{D}\Big)\ba{c}+(A_1^+\lra
A_1^-)\\ +(A_2^+\lra
A_2^-)\\-(e^{i\gamma}\lra e^{-i\gamma})\ea\right\}$$
is the homogeneous function in \eref{IvW}, read off from
\eref{IaFtotal}. Note that the factors $\frac1{\tan(\gamma)}\sim\cos(\gamma)$ cancel the apparent
singularity of the denominator $\frac1{(e_1e_2)}$ in \eref{IvW} when $\gamma=\frac\pi2$.

Using $\frac d{dz}\Li_2(z) =
-\frac{\log(1-z)}z$, and hence $\pa_{u}\Li_2(e^{\pm i\gamma}F(u)) = -\log\big(1-e^{\pm
  i\gamma}F(u)\big)\cdot \pa_u \log(F(u))$, it suffices to notice that
for all combinations of the signs
$$\pa_{u_1}\log\Big(\frac{A_1^{\pm_1} A_2^{\pm_2}}D\Big) = \frac{\sin(\gamma)}D\cdot
K_{\pm_1\pm_2}, \qquad K_{\pm_1\pm_2} := \pm_1 \frac{u_1u_2+v}{\sqrt{1-u_1^2}}
\pm_2\sqrt{1-u_2^2}.$$
Hence
$$v\pa_{u_1}H(u_1,u_2,v) = \frac12\frac\gamma{\tan(\gamma)}\pa_{u_1}\log(D) +
\frac{\pi}{2D}K_{++} + \frac
i{4D}\Big\{\sum_{\pm_1,\pm_2}K_{\pm_1\pm_2}\log\Big(1-e^{i\gamma}\frac{A_1^{\pm_1} A_2^{\pm_2}}{D}\Big)-(e^{i\gamma}\lra e^{-i\gamma})\Big\}.
$$
The first term equals the first term in \eref{paH} (for $k=1$).

Collecting all the resulting logarithms from the derivatives of the
dilog sum in
\eref{H}, and exploiting \eref{ApAm}, confirms after a lengthy
computation with ordinary logarithms the desired result \eref{paH} (Felix Tippner).

\newpage

{\bf Derivatives II}

\bfr
{\bf Lemma:} For $F(x)=\frac 1{x^2}$, and $x,e_1,e_2$ purely spatial, it holds
\bea{deri}\pa_x \lim_{a\to\infty} (I_{e_2}^aI_{e_1}F)(x)\equiv
(I_{e_2}I_{e_1}\pa F)(x) = \frac 12 \big[ f(e_1,e_2)\pa_x +
f(x,e_2)\pa_{e_1} + f(x,e_1)\pa_{e_2}\big] \log
\det{}_{xe_1e_2}.\quad
\eea
\efr
The differential identities in I immediately follow from this.

{\em Proof:} The total symmetry in the three vectors is expected from
the change of integration variables $t_1=s_1\inv$, $t_2= s_2s_1\inv$ in
$$(I_{e_2}I_{e_1}\pa F)(x) = -2\ioi ds_2\ioi ds_1 \frac
{x+s_1e_1+s_2e_2}{[(x+s_1e_1+s_2e_2)^2]^2}.$$
To prove the actual formula, we need some preparations.

Claim 1: For $\zeta_i$ defined as above,
$$\pa_x \Li_2\big(e^{i\gamma} e^{\pm(\zeta_1+\zeta_2)}\big) =
\mp\log\big(1-e^{i\gamma}
e^{\pm(\zeta_1+\zeta_2)}\big)\pa_x(\zeta_1+\zeta_2), $$
$$\pa_x \Li_2\big(-e^{i\gamma} e^{\pm(\zeta_1-\zeta_2)}\big) =
\mp\log\big(1+e^{i\gamma}
e^{\pm(\zeta_1-\zeta_2)}\big)\pa_x(\zeta_1-\zeta_2),$$
and the same with $e^{i\gamma}$ replaced by $e^{-i\gamma}$.

Claim 2: $$(1-e^{-i\gamma}e^{\zeta_1+\zeta_2})(1+e^{i\gamma}e^{-\zeta_1+\zeta_2})
= \frac {2\sin^2\gamma e^{\zeta_2}}{\sqrt D}\Big[-u_1+i\sqrt{1-u_1^2}\Big],$$
$$(1-e^{+i\gamma}e^{-\zeta_1-\zeta_2})(1+e^{-i\gamma}e^{\zeta_1-\zeta_2})
= \frac {2\sin^2\gamma e^{-\zeta_2}}{\sqrt D}\Big[u_1-i\sqrt{1-u_1^2}\Big],$$
and the same with $\zeta_1\lra \zeta_2$, and with $e^{i\gamma}\lra
e^{-i\gamma}$.

Claim 3: For $j=1,2$, $k=1,2$
$$\pa_x \zeta_j =
\frac{\sqrt{\det_{e_1e_2}}}{\sqrt{\det_{xe_j}}}\cdot \pa_{e_k}\log \det{}_{xe_1e_2}.$$

Claim 1 follows immediately from $\pa_z\Li_2(z) =
-\frac{\log(1-z)}z$. \qed

Claim 2 follows by a direct computation: multiply the brackets,
write $e^{\pm i\gamma} = \cos\gamma\pm i\sin\gamma$,
$e^{\zeta_j}=\frac {A_j^+}{\sqrt D}$, $e^{-\zeta_j}=-\frac
{A_j^-}{\sqrt D}$,
rearrange things such that $A_j^+\pm A_j^-$ appear, insert the
definitions of $A_j^\pm$, and finally use $v=-\cos\gamma$ and
$\sqrt{1-v^2}=\sin(\gamma)$. \qed

Claim 3:  First write $\zeta_j = \frac12 \log\frac{A_j^+}{-A_j^-}$ to
conclude
$$\pa_x \zeta_j =
\frac12\frac{A_j^+\pa_xA^-_j-A_j^-\pa_xA^+_j}D=\frac14\frac{(A_j^+-A_j^-)\pa_x(A_j^++A^-_j)-
  (A_j^++A_j^-)\pa_x(A_j^+-A^-_j)}D.$$
Insert the definitions to conclude
$$\pa_x \zeta_j 
=\frac12\frac{\sqrt{\det_{xe_j}\det_{e_1e_2}}\stackrel\lra\pa_x\big((xe_j)(e_je_k)-e_j^2(xe_k)\big)}{e_j^2\det_{xe_1e_2}}.$$
Work out the derivatives and recognize $\pa_{e_k}\det_{xe_1e_2}$ appearing
in the numerator. \qed

Now, in order to prove the Lemma, first notice that the derivative of
the first two terms equals the first term \eref{deri}, simply because
$\frac{\gamma}{\sin\gamma}=\sqrt{e_1^2e_2^2}f(e_1,e_2)$ and $\det_{xe_1e_2}=
x^2e_1^2e_2^2\cdot D$. Then use Claim 1 for each of the eight terms in the dilog sum. Collect
terms involving $\pa_x \zeta_1$ and $\pa_x \zeta_2$. Each of
them has a coefficient which is a logarithm of four factors in the
numerator and four factors in the denominator, that combine pairwise
to combinations as in Claim 2. Many positive factors cancel. The
result is
$$\pa_x \frac i{4\sqrt{\det_{e_1e_2}}}\big\{\hbox{dilog
  sum}\big\} = \frac i{4\sqrt{\det_{e_1e_2}}}
\sum_{j=1,2}\Big(\log\frac{-(xe_j)+i\sqrt{\det_{xe_j}}}{-(xe_j)-i\sqrt{\det_{xe_j}}}+
\log\frac{(xe_j)-i\sqrt{\det_{xe_j}}}{(xe_j)+i\sqrt{\det_{xe_j}}}\Big)\pa_x\zeta_j.$$
Use $\log\frac{a-ib}{a+ib} = \log\frac{-a+ib}{-a-ib}-2\pi i$ for $b>0$, so that
the bracket becomes
$$2\log\frac{-(xe_j)+i\sqrt{\det_{xe_j}}}{-(xe_j)-i\sqrt{\det_{xe_j}}}
-2\pi i = 4i\sqrt{\det{}_{xe_j}}f(x,e_j) -2\pi i.$$
The contribution from the subtractions:
$$\frac i{4\sqrt{\det_{e_1e_2}}}
\sum_{j=1,2} (-2\pi i\, \pa_x\zeta_j) =
\frac\pi{2\sqrt{\det_{e_1e_2}}}\cdot(\pa_x\zeta_1+\pa_x\zeta_2)$$
cancel against the contribution
$$\pa_x\Big(-\frac\pi{2\sin\gamma}\log\big(\frac{A_1^+A_2^+}D\big)\Big)
\equiv -\frac \pi{2\sqrt{\det_{e_1e_2}}}\cdot \pa_x \log
e^{\zeta_1+\zeta_2}$$
of the third term in \eref{IaFtotal}. 
Finally use Claim 3, to see that the remaining terms
$$\frac i{4\sqrt{\det_{e_1e_2}}}
\sum_{j=1,2} 4i\sqrt{\det{}_{xe_j}}f(x,e_j) \pa_x\zeta_j$$
coincide with the last two terms in \eref{deri}. \qed

{\bf Limit $\gamma\to 0$}

A generic configuration to study the limit $e_1\to e_2$ $\Leftrightarrow$
$\gamma\to0$ in the purely spatial case is
$$e_1=\bpm 0\\ \sin\frac\gamma2 \\ 0 \\ \cos\frac\gamma2 \epm,
\qquad e_2 =\bpm 0\\ -\sin\frac\gamma2 \\ 0 \\ \cos\frac\gamma2
\epm,\qquad x=r\bpm 0\\ n_1\\ n_2\\ n_3\epm \quad[\hbox{or}\,
=\rho\bpm n_2\sinh\tau\\ n_1\\n_2\cosh\tau\\n_3\epm]$$
with
$$u_1=-n_3\cos\frac\gamma2 -n_1\sin\frac\gamma2, \qquad u_2=-n_3\cos\frac\gamma2 +n_1\sin\frac\gamma2,$$
$$v=-\cos\gamma, \qquad \sqrt{1-v^2}=\sin\gamma,$$
$$D=n_2^2\cdot\sin^2\gamma,\qquad A_k^\pm = \pm e^{\pm\zeta_k}\cdot\vert n_2\vert\sin\gamma,$$
$$e^{\pm\zeta_1} = \frac{\sqrt{1-(n_1\sin\frac\gamma2+n_3\cos\frac\gamma2)^2}\pm n_1\cos\frac\gamma2\mp n_3\sin\frac\gamma2}{\vert n_2\vert},\quad
e^{\pm\zeta_2} = \hbox{the same with $n_1\to-n_1$}.$$

We compare the limit of \eref{IaFtotal} with the exact result at $e_1=e_2(=e)$, 
\eref{e-e'}. The latter has no other singularity than the logarithmic
IR divergence $-\frac12\log(\frac{(ae)^2}{x^2})$, that is identically present in
\eref{IaFtotal}. But various terms in \eref{IaFtotal} have singularities
$\sim\log(\gamma)$ that are not present in \eref{e-e'}, and therefore must cancel
each other. 

To compute \eref{IaFtotal} in the limit and compare with the exact result
\eref{e-e'}, notice that in the above configuration, we have for 
$\gamma\to0$,
$$e^{\zeta_2+\zeta_1}= 1-\frac{n_3}{\sqrt{1-n_3^2}}\cdot\gamma+
O(\gamma^2)  =e^{-a\gamma},\quad\hbox{where}\quad a=\frac{n_3}{\sqrt{1-n_3^2}}+O(\gamma),$$
$$e^{\zeta_2-\zeta_1}=:b=\frac{\sqrt{n_1^2+n_2^2}-n_1}{\sqrt{n_1^2+n_2^2}+n_1}+O(\gamma).$$
Thus, the dilog sum is
$$2i\IM
\big\{\Li_2\big(e^{(i-a)\gamma}\big)+\Li_2\big(e^{(i+a)\gamma}\big)+\Li_2\big(-e^{i\gamma}b\big)+\Li_2\big(-e^{i\gamma}b\inv\big)\big\}.$$
The arguments of the first two terms lie close to $1$ where the branch
cut starts.
\begin{itemize} \item
We compute the first two terms of the dilog sum: 
$$
2i\IM\Big\{\Li_2\big(e^{(i-a)\gamma}\big)+\Li_2\big(e^{(i+a)\gamma}\big)\Big\}=$$
$$
=2i\IM\Big\{-\Li_2\big(1-e^{(i-a)\gamma}\big)-(i-a)\gamma\log(1-e^{(i-a)\gamma})-\Li_2\big(1-e^{(i+a)\gamma}\big)-(i+a)\gamma\log(1-e^{(i+a)\gamma})\Big\}=
$$
$$
=2i\IM\Big\{(i-a)\gamma-(i-a)\gamma\log\big(-(i-a)\gamma\big)+(i+a)\gamma-(i+a)\gamma\log\big(-(i+a)\gamma\big)\Big\}+O(\gamma^2)=
$$
$$
=2i\IM\gamma\cdot\big\{2-(i-a)\log\big(-(i-a)\gamma\big)-(i+a)\log\big(-(i+a)\gamma\big)\big\}+O(\gamma^2)=
$$
$$
=2i\IM\gamma\cdot\big\{2-2\log(\gamma)+(a-i)\log(a-i)-(a+i)\log\big(-a-i\big)\big\}+O(\gamma^2)=
$$
\bea{Lising}
=2i\gamma \cdot \big\{2-2\log(\gamma)-\log(a^2+1) -a(2\alpha-\pi)\big\} +O(\gamma^2),
\eea
where $a+i=e^{i\alpha}\sqrt{a^2+1}$, ie, $n_3=\cos(\alpha)$.
\item We compute the last two terms of the dilog sum: Because the
  arguments lie close to the negative real
axis, the imaginary part in first order can be computed with the
derivative  $\Li_2'(z)=-\frac{\log(1-z)}z$: 
  $$2i\IM\Li_2\big(-e^{i\gamma} b\big) + (b\lra b^{-1}) =
  2\gamma\cdot\pa_\gamma \Li_2\big(-e^{i\gamma} b\big) + (b\lra b^{-1})+O(\gamma^2)=$$
$$=2i\gamma \cdot (-b)\Li_2'(-b)+
(b\lra b^{-1})+O(\gamma^2)=-2i\gamma\cdot\log\big((1+b)(1+
b^{-1})\big)+O(\gamma^2)=
$$
$$=-2i\gamma\cdot\log\frac{4(1-n_3^2)}{n_2^2}+O(\gamma^2).$$
\item The other terms in \eref{IaFtotal} near $\gamma=0$ are
  $$ -\frac12\log\big(\frac{a^2e^2}{x^2}\big)-
  \log\frac{\gamma}{\vert n_2\vert} - \frac\pi{2\gamma}\log
  e^{-a\gamma}+O(\gamma).$$
  Thus, near $\gamma=0$, \eref{IaFtotal} without the IR singularity $-\frac12\log\big(\frac{a^2e^2}{x^2}\big)$,
  equals 
$$-  \log\frac{\gamma}{\vert n_2\vert} +\frac\pi2\cdot a-\frac
  12\big\{2-2\log\gamma-\log(a^2+1) -a(2\alpha-\pi)\big\} +
  \frac12\log\frac{4(1-n_3^2)}{n_2^2}+O(\gamma)=$$
  $$= -1+\frac12\log(a^2+1) +a\alpha +
  \frac12\log(4(1-n_3^2))+O(\gamma)=\log2-1+a\alpha.$$
In particular, the singular term $\log\gamma$ cancels. \end{itemize}

By comparison with \eref{e-e'}, this should
equal
$$\frac12\log4 + \frac1{e^2}+\frac{(xe)}{e^2}f(x,e) =\log2-1+
\frac1{2i}
\frac{n_3}{\sqrt{1-n_3^2}}\log\frac{n_3+i\sqrt{1-n_3^2}}{n_3-i\sqrt{1-n_3^2}}=\log2-1+\frac a{2i}\log\frac{a+i}{a-i}.$$
This is easily seen to be true. \qed

\newpage

{\bf Limit $\gamma\to \pi$}

A generic configuration to study the limit $e_1\to -e_2$ $\Leftrightarrow$
$\theta=\pi-\gamma\to0$ in the purely spatial case is
$$e_1=\bpm 0\\ \sin\frac\theta2 \\ 0 \\ \cos\frac\theta2 \epm,
\qquad e_2 =\bpm 0\\ \sin\frac\theta2 \\ 0 \\ -\cos\frac\theta2
\epm,\qquad x=r\bpm 0\\ n_1\\ n_2\\ n_3\epm$$
with
$$u_1=-n_3\cos\frac\theta2 -n_1\sin\frac\theta2, \qquad u_2=n_3\cos\frac\theta2 -n_1\sin\frac\theta2,$$
$$v=\cos\theta, \qquad \sqrt{1-v^2}=\sin\theta,$$
$$D=n_2^2\cdot\sin^2\theta,\qquad A_k^\pm = \pm e^{\pm\zeta_k}\cdot\vert n_2\vert\sin\gamma,$$
$$e^{\pm\zeta_1} = \frac{\sqrt{1-(n_1\sin\frac\theta2+n_3\cos\frac\theta2)^2}\mp n_1\cos\frac\theta2\pm n_3\sin\frac\theta2}{\vert n_2\vert},\quad
e^{\pm\zeta_2} = \hbox{the same with $n_3\to-n_3$}.$$
In this case,
$$e^{\zeta_2-\zeta_1}=e^{-a\theta}\quad\hbox{where}\quad
a=\frac{n_3}{\sqrt{1-n_3^2}}+O(\theta), \quad\hbox{and}\quad
e^{\zeta_2+\zeta_1}=b
=\frac{\sqrt{n_1^2+n_2^2}-n_1}{\sqrt{n_1^2+n_2^2}+n_1} + O(\theta),$$
and because $e^{i\gamma}=-e^{-i\theta}$, the dilog sum becomes
$$-2i\IM
\big\{\Li_2\big(-e^{i\theta}b\big)+\Li_2\big(-e^{i\theta}b\inv\big)+\Li_2\big(e^{(i-a)\theta}\big)+\Li_2\big(e^{(i+a)\theta}\big)\big\},$$
where now the last two terms contribute to the singularity. But the
sum is the same as in the previous limit $\gamma\to0$, with $\gamma$ replaced by
$\theta$ and an overall sign.
Proceeding as before, we find that there survive integrable singularities
$\log\theta/\theta$ and $1/\theta$ as well as finite terms:
$$\frac
1{2\theta}\Big[(\pi-\theta)\log(\mu_v^2x^2)-2\pi\log\theta+\pi\log
n_2^2 - \pi\log b \Big] +a\big(i\log\frac{a+i}{a-i}+\pi\big)$$ 
plus a constant: $1-\frac12\log(n_2^2(1+a^2)(1+b)(1+b\inv))=1-\log
2$. 

Christian has redone the computation with $x_0\neq 0$. He found the
same result with $n_2^2$ replaced by $\frac{x_2^2-x_0^2}{r^2-x_0^2}$, 
$a=\frac{n_3}{\sqrt{n_1^2+n_2^2}}$ replaced by $\wt a = \frac{x_3}{\sqrt{x_1^2+x_2^2-x_0^2}}$ and
$b=\frac{\sqrt{n_1^2+n_2^2}-n_1}{\sqrt{n_1^2+n_2^2}+n_1}$ replaced by
$\wt b=
\frac{\sqrt{x_1^2+x_2^2-x_0^2}-x_1}{\sqrt{x_1^2+x_2^2-x_0^2}+x_1}$. There
occurs the same cancellation of finite terms
$1-\frac12\log\Big(\frac{x_2^2-t^2}{r^2-t^2}(1+\wt a^2)(1+\wt b)(1+\wt b\inv)\Big)=1-\log 2$.

Recall that $\mu_v^2$ may
depend on $\theta$ (so that it may cancel the $\log\theta$ term), but not on $\vec n$.

For the two-point fn $\erw{\phi(x,e)\phi(x',e')}$, this expression should be
multiplied by $-(ee') =(e_1e_2)=\cos\theta=1+O(\theta^2)$. If integrated with
smearing functions $c$ as in \sref{s:vertex}, $\theta$ is the angle
between $\vec e=\vec e_1$ and $\vec e\,'=-\vec e_2$ both in
the support of $c$, so that for $c$ with small support, the
denominator $1/\theta$ makes $\erw{c,c}$ increase. But one should
be aware that the components $n_i$ of the direction of $\vec x$ must be taken in
a frame depending on $\vec
e_1$ and $\vec e_2$, in which these vectors have the standard form as
above, with zero $2$-components. In particular, ``$n_1$'' and
``$n_3$'' in the above must be
understood as the
components of $\vec x/r$ in the direction of $\vec e_1+\vec e_2$ resp
$\vec e_1-\vec e_2$, and ``$n_2$'' as the component perpendicular to $\vec e_1$ and $\vec e_2$.
Thus, the components $n_i$ will be (discontinuous near $\vec e_1=-\vec e_2$!) functions of $\vec e_1$ and $\vec
e_2$, which must be taken into account when smearing. This may be
tricky, even (or especially) when the support is small.

The exponential of the smeared two-point function (multiplying the Dirac two-point
function) is needed for the spectral analysis (dissolution of the
mass-shell) of the states
$\psi_c(f)^*\Omega$ 

More precisely, I need this formula also for singular configurations,
in particular for $x^0\neq0$\dots


\end{document}